\documentclass[aps,prb,twocolumn,10pt,a4paper,showpacs]{revtex4-1}
\usepackage{graphicx}
\usepackage{amsmath}
\usepackage{amssymb}
\usepackage{bm}

\begin{document}


\newcommand{\vt}{\tilde V}
\newcommand{\tk}{T_\mathrm{K}}
\newcommand{\boldk}{{\boldsymbol{\mathrm{k}}}}
\newcommand{\K}{\boldk}
\newcommand{\modmu}{|\mu|}
\newcommand{\hfmu}{|\mu_0|}
\newcommand{\bra}[1]{\langle #1 |}
\newcommand{\ket}[1]{| #1 \rangle}
\newcommand{\cre}[1]{c_{#1}^\dagger}
\newcommand{\ann}[1]{c_{#1}^{\phantom{\dagger}}}
\newcommand{\dbyd}[2]{\left(\frac{\partial #1}{\partial #2}\right)}
\newcommand{\sgn}{\operatorname{sgn}}
\newcommand{\E}{\mathrm{e}}
\newcommand{\D}{\;\mathrm{d}}
\newcommand{\I}{\mathrm{i}}
\newcommand{\ex}[1]{\langle #1 \rangle}
\newcommand{\tr}{\operatorname{Tr}}
\newcommand{\brapn}{\bra{\Psi_0^N}}
\newcommand{\ketpn}{\ket{\Psi_0^N}}
\newcommand{\deltar}{\Delta_\mathrm{R}}
\newcommand{\deltai}{\Delta_\mathrm{I}}
\newcommand{\deltao}{\Delta_0}
\newcommand{\sigr}{\Sigma^\mathrm{R}}
\newcommand{\sigi}{\Sigma^\mathrm{I}}
\newcommand{\gupa}{G_\uparrow^\mathrm{A}(\omega)}
\newcommand{\gupb}{G_\uparrow^\mathrm{B}(\omega)}
\newcommand{\gdna}{G_\downarrow^\mathrm{A}(\omega)}
\newcommand{\gdnb}{G_\downarrow^\mathrm{B}(\omega)}
\newcommand{\ga}[1]{G_{#1}^\mathrm{A}(\omega)}
\newcommand{\gb}[1]{G_{#1}^\mathrm{B}(\omega)}
\newcommand{\half}{\frac{1}{2}}
\newcommand{\niup}{\hat n_{\mathrm{i}\uparrow}}
\newcommand{\nidn}{\hat n_{\mathrm{i}\downarrow}}
\newcommand{\nisig}{\hat n_{\mathrm{i}\sigma}}
\newcommand{\scrgu}{\mathcal{G}_\uparrow}
\newcommand{\scrgd}{\mathcal{G}_\downarrow}
\newcommand{\ut}{\tilde U}
\newcommand{\dc}{\tilde \delta_\mathrm{c}}
\newcommand{\dt}{\tilde \delta}
\newcommand{\ot}{\tilde \omega}
\newcommand{\omegam}{\omega_\mathrm{m}}
\newcommand{\hc}{\hat{H}_\mathrm{c}}
\newcommand{\hh}{\hat{H}_\mathrm{h}}
\newcommand{\hp}{\hat{H}_\mathrm{p}}
\newcommand{\mub}{\mu_\mathrm{B}}
\newcommand{\epsi}{\epsilon_\I}
\newcommand{\epsit}{\tilde\epsi}
\newcommand{\simutoinf}{\overset{U\to\infty}{\sim}}
\newcommand{\simuttoinf}{\overset{\tilde U\to\infty}{\sim}}
\newcommand{\omo}{\ot^0_\mathrm{m}}
\newcommand{\om}{\omega_\mathrm{m}}
\newcommand{\eltup}{$| \epsilon | < U' $}
\newcommand{\egtup}{$| \epsilon | > U' $}
\newcommand{\eequp}{$| \epsilon | = U' $}
\newcommand{\eeqz}{$| \epsilon | = 0 $}
\newcommand{\eph}{$| \epsilon | = U' + \dfrac{1}{2} U $}
\newcommand{\gt}{\tilde\Gamma}
\newcommand{\jt}{\tilde J}
\newcommand{\jpt}{\tilde J'}
\newcommand{\ua}{\uparrow}
\newcommand{\da}{\downarrow}
\newcommand{\et}{\tilde \epsilon}
\newcommand{\gte}{\tilde \Gamma = \frac{\Gamma}{D} =}
\newcommand{\ute}{\tilde U = \frac{U}{\pi \Gamma} =}
\newcommand{\jte}{\tilde J = \frac{J}{D} =}
\newcommand{\jpte}{\tilde J' = \frac{J'}{D} =}
\newcommand{\phs}{\epsilon=-\frac{U}{2}}
\newcommand{\thalf}{\tfrac{1}{2}}
\newcommand{\jpcte}{\tilde J'_{c}=\frac{J'_{c}}{D}=}
\newcommand{\ect}{\tilde \epsilon_{c}=\frac{\epsilon_{c}}{\pi \Gamma}=}
\newcommand{\nt}{\langle n_{2} \rangle}
\newcommand{\delt}{J-J'+\half \delta}
\newcommand{\delts}{\sqrt{(J-J'+\half \delta)^{2}+\frac{3}{4}\delta^{2}}}
\newcommand{\ttt}{\tilde t = \frac{t}{\pi \Gamma}}
\newcommand{\tpt}{\tilde t' = \frac{t'}{\pi \Gamma}}


\newcommand{\comment}[1]{}

\newcommand{\w}{\omega}
\newcommand{\wpr}{\omega^{\prime}}

\newcommand{\Ss}{\Sigma_{\sigma}}
\newcommand{\Ssp}{\Sigma_{\sigma}^{\prime}}
\newcommand{\SsR}{\Sigma_{\sigma}^{R}}
\newcommand{\SsI}{\Sigma_{\sigma}^{I}}
\newcommand{\Gs}{G_{\sigma}}

\newcommand{\tI}{\tilde{I}_{\sigma\sigma^{\prime}}}
\newcommand{\tgamma}{\tilde{\Gamma}_{\sigma\sigma^{\prime}}}
\newcommand{\pd}{\phantom\dagger}
\newcommand{\chic}{\chi^{\pd}_{\mathrm{c,imp}}}
\newcommand{\chis}{\chi^{\pd}_{\mathrm{s,imp}}}
\newcommand{\tm}{\tilde{\mu}}
\newcommand{\chisloc}{\chi^{\pd}_{s}}
\newcommand{\nimp}{n_{\mathrm{imp}}}
\newcommand{\nimps}{n_{\mathrm{imp},\sigma}}
\newcommand{\nimpas}{n_{\mathrm{imp},A\sigma}}
\newcommand{\mimp}{m_{\mathrm{imp}}}
\newcommand{\mimpa}{m_{\mathrm{imp},A}}
\newcommand{\rhoimps}{\Delta\rho_{\mathrm{imp},\sigma}(\w, h)}

\newcommand{\Gc}{G_{c}(\w)}
\newcommand{\gc}{g_{c}(\w)}
\newcommand{\gctil}{\tilde{g}_{c}(\w)}
\newcommand{\Gf}{G_{f}(\w)}
\newcommand{\Sigf}{\Sigma_{f}(\w)}
\newcommand{\teff}{t_{*\mathrm{eff}}}
\newcommand{\wtil}{\tilde{\w}}
\newcommand{\neff}{\eta_{\mathrm{eff}}}
\newcommand{\mutil}{|\tilde{\mu}|}


\title{Mott transitions in the Periodic Anderson Model}


\author{David E. Logan, Martin R. Galpin and Jonathan Mannouch}
\affiliation{Department of Chemistry, Physical and Theoretical Chemistry, Oxford University, South Parks Road, Oxford, OX1 3QZ, United Kingdom}

\date{\today}

\begin{abstract}
The periodic Anderson model (PAM) is studied within the framework of dynamical mean-field theory, 
with particular emphasis on the interaction-driven Mott transition it contains, and on 
resultant Mott insulators of both Mott-Hubbard and charge-transfer type. The form of the PAM phase diagram 
is first deduced on general grounds using two exact results, over the full range of model parameters and
including metallic, Mott,  Kondo and band insulator phases.
The effective low-energy model which describes the PAM in the vicinity of a
Mott transition is then shown to be a one-band Hubbard model, with effective hoppings that are not  
in general solely nearest neighbour, but decay exponentially with distance. This mapping is shown to have a 
range of implications for the physics of the problem, from phase boundaries to single-particle dynamics; 
all of which are confirmed and supplemented by NRG calculations.
Finally we consider the locally degenerate, non-Fermi liquid Mott insulator, to describe
which requires a two-self-energy description. This is shown to yield a number of exact results for the
associated local moment, charge, and interaction-renormalised levels, together with a generalisation of 
Luttinger's theorem to the Mott insulator.
\end{abstract}


\maketitle

\section{Introduction}
\label{section:intro}

The periodic Anderson model (PAM) is one of the classic models of strongly correlated electron systems.~\cite{hewsonbook}
Minimalist by design, but physically richer than the canonical one-band Hubbard model, the PAM 
is a two-band model with the two orbitals per lattice site coupled by a local one-electron hybridization. One
band is localised but correlated, with on-level interaction $U$, the other uncorrelated
but itinerant. While traditionally considered in the context of $f$-electron systems~\cite{hewsonbook}
-- heavy fermion materials -- the correlated orbitals can equally refer to the localised $d$-orbitals
of transition metal oxide and related materials, enabling access also to the basic physics of these systems.

Within the powerful framework of dynamical mean-field theory (DMFT),~\cite{MetznerVollhardt1989,MuellerH1989a,*MuellerH1989b,GeorgesKotliarPRB1992,JarrellPRL1992,dmftkotliar,PruschkeJarrellFreericksAdvPhys1995} 
which is formally exact in the limit of large coordination number and known to capture well many properties of real 
materials,~\cite{KotVollPhyToday2004,* KotliarEtAlRMP2006} numerous facets of the PAM have been 
extensively studied over many years.~\cite{dmftkotliar,PruschkeJarrellFreericksAdvPhys1995,CzychollPRL92,JarrellPAMPRL93,*JarrellPAMPRB95,SunPAMPRB93,GreweNCA88,*GreweNCA89,RozenbergEDPRB95,RozenbergIPT96,JarrellFreericksPRB97,*JarrellFreericksPRL98,*JarrellFreericksPRB99,TPRBJarrellPAMPRB2000,NSVMJHRKEPL2000,BurdinGeorgesPRL2000,vickiSPAMEPJB2003,*KIPAMJPCM2003,*ABG+NSV2007,RajaEPJB,rajapamtheory,*rajapamexp,FBACzycholTP_PAMtransportPRB2006,RajaEPL2007,SordiPRL2007,*SordiPRB2009,AmaricciPRL2008,Parihari+NSV2008,Burdin+VZPRB2009,TP+MVQPI_PRB2011,NSV+KumarJPCM2011,AmaricciPRB2012,PruschkeZitkoPRB2015}
The model is known to contains a diverse range of phases,~\cite{dmftkotliar,PruschkeJarrellFreericksAdvPhys1995}  
including the metal, Kondo and band insulators -- all Fermi liquids since they are adiabatically connected to the non-interacting limit -- as well as a  Mott insulator phase and hence an underlying Mott metal-insulator 
transition.~\cite{SordiPRL2007,*SordiPRB2009}
Due to the two-band nature of the model, the resultant Mott phase also exemplifies
the  Zaanen-Sawatzky-Allen (ZSA) scheme,~\cite{Zaanen-S-A1985} distinguishing 
between Mott-Hubbard or charge-transfer insulators according respectively to whether $\Delta \gtrsim U$ or
$\Delta\lesssim U$, 
with $\Delta$ the charge transfer energy -- the one-electron energy gap between the  uncorrelated and correlated local levels.

In the present paper we report a detailed DMFT study of the PAM, with particular emphasis on understanding three related aspects of it: 
(a) its phase diagram over essentially the full range of underlying model parameters; (b) the connection between the Mott transition in the PAM and that occurring in a one-band Hubbard model; and
(c) the non-Fermi liquid Mott insulator phase itself.
Since the paper is quite wide ranging we begin with an overview of it, and some of the issues 
to be addressed.


\subsection{Overview}
\label{subsection:overview}

The model and DMFT background is first summarised in sec.\ \ref{section:model}. 
The basic parameter space for the PAM is of course large compared e.g.\ to a nearest neighbour Hubbard model.
 In addition to the one-electron hopping $t_{*}$, which connects and broadens the uncorrelated conduction band levels (with level energies $\epsilon_{c}$), we 
take its `bare' parameters to be $\epsilon_{c}, U, V$ and $\eta_{f}$; with $U$ the local Coulomb repulsion for the correlated levels with energies $\epsilon_{f}$, and $V$ the one-electron hybridization coupling local $c$- and 
$f$-levels (for specificity we refer to the correlated orbitals as $f$-levels).  As for an Anderson impurity model, the asymmetry of the correlated level is embodied in~\cite{anderson,RajaEPJB}
$\eta_{f}=1+2\epsilon_{f}/U$, which controls whether the level is in a local moment, mixed-valent or empty-orbital regime.

Sec.\ \ref{section:MVsI} begins with two exact results which play an important role in the subsequent analysis.  
 We first identify a general inequality that must be satisfied by any insulator -- Mott, Kondo or band -- 
relating $\epsilon_{c}$ and  $V$ to the interaction-renormalised $f$-level energy $\epsilon_{f}^{*}$.
The second result relates to Fermi liquid phases, encompassing metallic, Kondo and band insulator phases.
It is has been known for some time,~\cite{RajaEPJB} and relates the total charge per site
to $\epsilon_{c}, V$ and $\epsilon_{f}^{*}$. Conjoined with a series of simple physical arguments, these results are 
shown in secs.\ \ref{subsection: Phasediag}, \ref{subsection: Phasediagtake2} to enable the general form and structure of the PAM phase diagram to be deduced, as a function of $\epsilon_{c}/t_{*}$, $U/t_{*}$ and 
$\eta_{f}$, encompassing all phases of the model.

Full numerical renormalization group (NRG) calculations of the phase boundaries are also given. They confirm the 
 deductions made and, for the Mott transition in particular, are seen to provide  clear examples of the ZSA 
classification scheme.~\cite{Zaanen-S-A1985} 

The resultant behaviour nonetheless also raises a number of questions.
For example, on progressively depleting the conduction band occupancy by raising its center of gravity relative to the Fermi level  -- by ramping up $\epsilon_{c}/t_{*}\gg 1$ -- the critical $U/t_{*}$ for the Mott transition asymptotically vanishes; which begs an explanation, given that the Mott transition is a paradigm of strong correlations. More generally, the question 
naturally arises~\cite{SordiPRL2007,*SordiPRB2009} as to what extent the Mott transition in the PAM
is similar to that occurring in a simpler one-band Hubbard model. We consider these and other basic matters in the remainder of the paper.

While it has previously been thought that the low-energy properties of the PAM close to a Mott transition
cannot be described generally in terms of a one-band Hubbard model,~\cite{SordiPRL2007,*SordiPRB2009,AmaricciPRB2012} 
one central result of the present work is to show (sec.\ \ref{section:Hubbardmapping})
that the effective low-energy model which describes the PAM in the vicinity of the Mott transition is in fact precisely a single-band Hubbard model. The one-electron hoppings in this effective Hubbard model 
are in general long-ranged, decaying exponentially with the topological distance between sites 
(providing a connection to models hitherto studied in the non-interacting, tight-binding 
limit~\cite{Eckstein2005}); and with an effective range found to be controlled by $\epsilon_{c}/t_{*}$.
For the particular regime $\epsilon_{c}/t_{*} \gg 1$  -- relevant close to the transition to a Mott-Hubbard insulator in
the ZSA scheme~\cite{Zaanen-S-A1985} -- the effective hopping, $\teff$, is in practice nearest neighbour (NN) only, but
with $\teff$  strongly reduced from the bare PAM hopping $t_{*}$; which 
underlies the narrowness of  electron bands  anticipated in the vicinity of the transition to Mott-Hubbard 
insulators.~\cite{Zaanen-S-A1985} Away from this asymptotic regime however, and in particular close to the Mott transition to charge-transfer insulators, the long-ranged nature of the hoppings is central to the problem.

In sec.\  \ref{subsection:qualitativeobs} we also touch briefly on the Kondo lattice model (KLM),
pointing out that, while commonly used as a proxy model in regimes where the correlated levels are almost singly occupied, 
the rich physics of Mott transitions in the PAM is simply absent in the KLM.

Given the mapping to an effective Hubbard model, we revisit the phase diagram
in sec.\ \ref{section:pdexplaining}. That the effective model is naturally specified by fewer parameters than the PAM, is first shown on general grounds to imply a scaling collapse of phase boundaries for the Mott transition (whereby the hybridization $V$ effectively scales out of the problem); as is indeed verified
in NRG calculations. Physical arguments are then used to obtain simple analytical estimates
for the transition to a Mott insulator -- including the approach to it from both the electron- and hole-doped sides 
(in regimes where both are possible) -- as well as that for the metal to band insulator transitions.
These too are found to agree well with NRG results for the full PAM.

 Single-particle dynamics in the metal close to the Mott transition are considered in 
sec.\ \ref{section:spdynamics}, to exemplify typical behaviour associated with Mott transitions to both 
Mott-Hubbard and charge-transfer insulators. The approach to the transition in either case is 
characterised by a vanishing low-energy scale, embodied in a low-energy Kondo resonance which narrows progressively 
as the transition is approached, and in terms of which the $f$- and $c$-electron spectra exhibit universal scaling. 
Scaling spectra are however found to be quite distinct in the Mott-Hubbard and charge-transfer regimes, 
as too are the associated bandwidth scales (`narrow' versus `broad' bands), and  the character of the insulating gaps.

Sec.\ \ref{section:MottPhase} is concerned with the paramagnetic Mott insulator phase  \emph{per se}. This 
presents a greater theoretical challenge than the metal, since the Mott insulator is not adiabatically connected to 
the non-interacting limit and as such is not a Fermi liquid. The electron spin degrees of freedom -- which in the metallic phase are quenched completely by the Kondo effect -- are incompletely quenched in the Mott insulator, resulting in a locally degenerate ground state with a residual local magnetic moment. To handle this degenerate Mott insulator requires a 
two-self-energy description.~\cite{LTG2014,DELMRG2016} This enables a number of exact results to be obtained
for the local moment, charge, and associated renormalised levels; which are also shown to be well captured by NRG 
calculations.

Finally, in sec.\ \ref{subsection:renormnormal} we consider the standard Luttinger 
integral $I_{L}$,~\cite{LW1960,*Luttingerf1960FermiSurf,*Luttingerf1961,LTG2014,DELMRG2016} the vanishing of which throughout
the metal is tantamount to Luttinger's theorem,~\cite{dmftkotliar} and reflects the Fermi liquid nature of that phase.
While this result naturally does not hold in the Mott insulator, a generalisation of it is proven,
showing that $I_{L}$ has constant but now non-zero magnitude throughout the entire Mott phase.
We have recently shown this result holds also for local moment phases of a wide range of quantum impurity 
models,~\cite{LTG2014} and for the one-band Hubbard model within DMFT;~\cite{DELMRG2016} suggesting its ubiquity as 
a hallmark of locally degenerate ground states, and indicative of their perturbative continuity to
the uncoupled atomic limit.


\section{Model and background}
\label{section:model}

The model  is given in standard notation by
\begin{subequations}
\label{eq:11}
\begin{align}
H~=~&H_{c}~+~H_{f}~+~H_{\mathrm{hyb}}
\label{eq:11a}
\\
H_{c}~=~&
~\sum_{i,\sigma}\epsilon_{c}^{\pd}
c_{i\sigma}^{\dagger}c_{i\sigma}^{\pd}
~-~
t\sum_{(i,j),\sigma}c_{i\sigma}^{\dagger}c_{j\sigma}^{\pd}
\label{eq:11b}
\\
H_{f}~=~& \sum_{i,\sigma}\left(
\epsilon_{f}^{\pd} +\tfrac{1}{2}U
f_{i-\sigma}^{\dagger}f_{i-\sigma}^{\pd}
\right)
f_{i\sigma}^{\dagger}f_{i\sigma}^{\pd}
\label{eq:11c}
\\
H_{\mathrm{hyb}}~=~& V\sum_{i,\sigma}\left(f_{i\sigma}^{\dagger}c_{i\sigma}^{\pd}+\mathrm{h.c.}\right)
\end{align}
\end{subequations}
with local $\sigma$-spin number operators $\hat{n}_{c\sigma} =c_{i\sigma}^{\dagger}c_{i\sigma}^{\pd}$ and
$\hat{n}_{f\sigma} =f_{i\sigma}^{\dagger}f_{i\sigma}^{\pd}$. $H_{c}$ represents the non-interacting conduction band.
Its nearest neighbour  hopping $t$ is rescaled within DMFT as~\cite{dmftkotliar,PruschkeJarrellFreericksAdvPhys1995} 
$t=t_{*}/(2\sqrt{Z_{c}})$ with coordination number $Z_{c} \rightarrow \infty$; 
and the only relevant property of the $c$-band energy dispersion is $\rho_{0}(\epsilon)$, the free ($V=0$) density of 
states (DoS) for $\epsilon_{c}=0$. This we take to be of standard compact, semicircular form
\begin{equation}
\label{eq:12}
\rho_{0}^{\pd}(\epsilon)~=~ \frac{2}{\pi t_{*}} \sqrt{ 1- (\epsilon/t_{*})^{2}}
\end{equation}
with band halfwidth $\tfrac{1}{2}W=t_{*}$, corresponding formally to a Bethe lattice.
$H_{f}$ specifies the correlated levels, while $H_{\mathrm{hyb}}$ hybridizes the $c$- and $f$-levels via a local one-electron matrix element $V$. 
While referring to the correlated orbitals as $f$-levels, we remind the reader that they could equally refer to
the $d$-levels of transition metal systems (with e.g.\ ligand $p$-levels for the conduction band).

With $t_{*}$ as the basic energy unit, the model is characterised by four `bare' parameters, $\epsilon_{c}/t_{*}$, 
$\epsilon_{f}/t_{*}$, $U/t_{*}$ and $V/t_{*}$.
Equivalently and more usefully, as employed in the following, it can  be parameterised by
 $\epsilon_{c}/t_{*}$, $U/t_{*}$, $V/t_{*}$ and $\eta_{f}$, with the $f$-level asymmetry $\eta_{f}$ given 
by~\cite{anderson,RajaEPJB}
\begin{equation}
\label{eq:13}
\eta_{f}^{\pd}~=~ 1 +\frac{2\epsilon_{f}}{U}~.
\end{equation}
For $\epsilon_{f}$ in the range $-U<\epsilon_{f} <0$ ($\equiv E_{F}$, the Fermi level), $\eta_{f}$ lies in the range 
$-1 <\eta_{f} <1$. Here an uncoupled $f$-level is singly occupied only, thus carrying a local moment. For obvious physical reasons this is often the regime of primary interest in the full system, although we will not restrict ourselves to it 
in part because (see secs.\ \ref{section:MVsI} \emph{ff}) Mott insulators are not confined to $|\eta_{f}| <1$.


\subsubsection{DMFT equations}
\label{subsubsection:DMFTeqs}

The local $c$- or $f$-electron charge, $n_{c}=\langle\hat{n}_{c}\rangle =\sum_{\sigma}\langle\hat{n}_{c\sigma}\rangle$ 
or $n_{f}=\langle\hat{n}_{f}\rangle$, is of course related to the retarded propagator
$G_{c}(\w)$ ($\leftrightarrow G_{c}(t)=-i\theta(t)\langle \{c_{i\sigma}^{\pd}(t),c_{i\sigma}^{\dagger}\}\rangle$)  
or $G_{f}(\w)$, by
\begin{equation}
\label{eq:14}
\tfrac{1}{2}n_{\nu}~=~-\tfrac{1}{\pi}\mathrm{Im}\int^{0}_{-\infty}d\w ~G_{\nu}(\w)~~~~:~\nu = c,f
\end{equation}
with $\w =0$ the Fermi level. The local propagators $G_{\nu}(\w) \equiv G_{\nu,ii;\sigma}(\w)$ are naturally 
independent of both site $i$ and spin $\sigma$ for the homogeneous paramagnetic phases we consider.

Within DMFT the propagator is given by~\cite{RajaEPJB}
\begin{equation}
\label{eq:15}
G_{\nu}(\w)~=~ \int^{\infty}_{-\infty}d\epsilon ~\rho_{0}(\epsilon) G_{\nu}(\epsilon; \w)
\end{equation}
with
\begin{equation}
\label{eq:16}
G_{c}(\epsilon;\w)~=~ \left[
\w^{+} -\epsilon_{c}^{\pd}- \frac{V^{2}}{\w^{+}-\epsilon_{f}^{\pd}  -\Sigma_{f}^{\pd}(\w)}-\epsilon
\right]^{-1}
\end{equation}
and
\begin{subequations}
\label{eq:17}
\begin{align}
&G_{f}(\epsilon;\w)= \left[
\w^{+} -\epsilon_{f}^{\pd}- \Sigma_{f}^{\pd}(\w)-\frac{V^{2}}{\w^{+}-\epsilon_{c}^{\pd}  -\epsilon}
\right]^{-1}
\label{eq:17a}
\\
=&~
\frac{1}{\w^{+}-\epsilon_{f}^{\pd}- \Sigma_{f}^{\pd}(\w)}
+\frac{V^{2}}{[\w^{+}-\epsilon_{f}^{\pd}- \Sigma_{f}^{\pd}(\w)]^{2}}G_{c}(\epsilon;\w)
\label{eq:17b}
\end{align}
\end{subequations}
where $\Sigma_{f}(\w) = \Sigma_{f}^{R}(\w)-i\Sigma_{f}^{I}(\w)$ is the $f$-level interaction self-energy (purely local, independent of $\epsilon$) and $\w^{+} = \w+i0+$.

With $\rho_{0}(\epsilon)$ as in eq.\ \ref{eq:12}, the $G_{\nu}(\w)$ may be written equivalently as
\begin{subequations}
\label{eq:18}
\begin{align}
G_{c}(\w)~=&~ \Big[
\w^{+} -\epsilon_{c}^{\pd}- \frac{V^{2}}{\w^{+}-\epsilon_{f}^{\pd}  -\Sigma_{f}^{\pd}(\w)}-\tfrac{1}{4}t_{*}^{2}G_{c}(\w)
\Big]^{-1}
\label{eq:18a}
\\
G_{f}(\w)=&~\Big[
\w^{+} -\epsilon_{f}^{\pd}- \Sigma_{f}^{\pd}(\w)-\frac{V^{2}}{\w^{+}-\epsilon_{c}^{\pd}  -\tfrac{1}{4}t_{*}^{2}G_{c}(\w)}
\Big]^{-1}
\label{eq:18b}
\\
\equiv&~\left[
\w^{+} -\epsilon_{f}^{\pd}- \Sigma_{f}^{\pd}(\w)-S_{f}^{\pd}(\w)
\right]^{-1}
\label{eq:18c}
\end{align}
\end{subequations}
with local $c$-level Feenberg self-energy~\cite{Feenberg,*Economou} $S_{c}(\w) =\tfrac{1}{4}t_{*}^{2}G_{c}(\w)$ 
(i.e.\ $S_{c}(\w) = \sum_{j}t^{2}G_{c,jj;\sigma}(\w)$ with sites $j$ NN to site $i$),
and local $f$-level  Feenberg self-energy $S_{f}(\w)=V^{2}[\w^{+}-\epsilon_{c} -\tfrac{1}{4}t_{*}^{2}G_{c}(\w)]^{-1}$.
Eq.\ \ref{eq:18} embodies the fact that within DMFT any lattice-fermion model reduces to a self-consistent quantum impurity 
model;~\cite{dmftkotliar,PruschkeJarrellFreericksAdvPhys1995} for it is precisely that for a 2-level Anderson impurity model in which the hybridization function coupling the $c$-level to the bath is $S_{c}(\w)$, while that coupling the $f$-level is 
$S_{f}(\w)$, and where the $S_{\nu}(\w)$ must be determined self-consistently. \\



We also touch here on particle-hole (ph) asymmetry, which enters the problem in two distinct ways: \\
(a) Conduction band asymmetry, embodied in 
$\epsilon_{c}$; which determines the centre of gravity of the free conduction band relative to the Fermi level,
as reflected in its density of states $d_{0}^{c}(\w) = \rho_{0}(\w -\epsilon_{c})$. Under a ph-transformation, $H_{c}$ itself (eq.\ \ref{eq:11b}) is ph-symmetric at the point $\epsilon_{c}=0$.\\
(b) $f$-level asymmetry, embodied in $\eta_{f}$ (eq.\ \ref{eq:13}), with $\eta_{f} =0$ at the ph-symmetric point of 
$H_{f}$ itself ($\epsilon_{f} =-U/2$).

Under a ph-transformation it is readily shown that  $H(\epsilon_{c},\eta_{f}) \equiv H(-\epsilon_{c},-\eta_{f})$ (for given 
$U,V$). Hence the full model is  ph-symmetric \emph{only} at the point $\epsilon_{c}=0=\eta_{f}$;
and only $\eta_{f} \geq 0$ need be considered. Likewise, with the $\epsilon_{c}, \eta_{f}$ dependence 
temporarily explicit, $\hat{n}_{\nu}(\epsilon_{c},\eta_{f})+\hat{n}_{\nu}(-\epsilon_{c},-\eta_{f})= 2$.
The total charge 
\begin{equation}
\label{eq:19}
n_{t}^{\pd}~:=~n_{c}{\pd}+n_{f}{\pd}
\end{equation}
thus satisfies
\begin{equation}
\label{eq:110}
n_{t}{\pd}(\epsilon_{c},\eta_{f})~+~n_{t}{\pd}(-\epsilon_{c},-\eta_{f})~=~ 4.
\end{equation}
For the self-energy, 
\begin{subequations}
\label{eq:111}
\begin{align}
\Sigma^{R}_{f}(\w; \epsilon_{c}, \eta_{f})~=&~U~-~\Sigma^{R}_{f}(-\w; -\epsilon_{c}, -\eta_{f})
\label{eq:111a}
\\
\Sigma^{I}_{f}(\w; \epsilon_{c}, \eta_{f})~=&~\Sigma^{I}_{f}(-\w; -\epsilon_{c}, -\eta_{f})
\label{eq:111b}
\end{align}
\end{subequations}
under a ph-transformation, as employed below.


\section{Metals \emph{vs} insulators}
\label{section:MVsI}

Leaving aside for the moment band insulators with $n_{t}=0$ or $4$, the PAM is well known to contain three distinct 
phases:~\cite{CzychollPRL92,JarrellPAMPRL93,*JarrellPAMPRB95,SunPAMPRB93,GreweNCA88,*GreweNCA89,RozenbergEDPRB95,RozenbergIPT96,JarrellFreericksPRB97,*JarrellFreericksPRL98,*JarrellFreericksPRB99,TPRBJarrellPAMPRB2000,NSVMJHRKEPL2000,BurdinGeorgesPRL2000,vickiSPAMEPJB2003,*KIPAMJPCM2003,*ABG+NSV2007,RajaEPJB,rajapamtheory,*rajapamexp,FBACzycholTP_PAMtransportPRB2006,RajaEPL2007,SordiPRL2007,*SordiPRB2009,AmaricciPRL2008,Parihari+NSV2008,Burdin+VZPRB2009,TP+MVQPI_PRB2011,NSV+KumarJPCM2011,AmaricciPRB2012,PruschkeZitkoPRB2015} 
a metal (M), a Kondo insulator (KI) and a Mott insulator (MI). Insulating states are characterised by a 
fixed total charge throughout the phase: $n_{t}=2$ for the KI and, for the MI,  $n_{t}=1$ or $3$.

Since metals \emph{vs} insulators are distinguished at the most basic level by the single-particle spectra 
$D_{\nu}(\w) = -\tfrac{1}{\pi}\mathrm{Im}G_{\nu}(\w)$ at the Fermi level, we first ask what can be inferred on 
general grounds for $D_{\nu}(\w =0)$.

The imaginary part of the self-energy at the Fermi level vanishes generically in all phases, 
$\Sigma_{f}^{I}(\w =0)=0$, reflecting the Fermi liquid character of the metal and the gapped nature of the insulators.
With this, the spectra $D_{\nu}(\w =0)$ follow simply from eqs.\ \ref{eq:15}-\ref{eq:17} as
\begin{subequations}
\label{eq:112}
\begin{align}
D_{c}(0)~=&~ \rho_{0}\Big(-\epsilon_{c}+\tfrac{V^{2}}{\epsilon_{f}^{*}}\Big)
\\
D_{f}(0)~=&~\frac{V^{2}}{{\epsilon_{f}^{*}}^{2}} ~\rho_{0}\Big(-\epsilon_{c}+\tfrac{V^{2}}{\epsilon_{f}^{*}}\Big).
\end{align}
\end{subequations}
Here,
\begin{equation}
\label{eq:113}
\epsilon_{f}^{*}~=~\epsilon_{f}~+~\Sigma^{R}_{f}(0)~=~ -\tfrac{U}{2}(1-\eta_{f})~+~\Sigma^{R}_{f}(0)
\end{equation}
is the interaction-renormalised $f$-level energy; with $\epsilon_{f}^{*}$ 
($\equiv \epsilon_{f}^{*}(\epsilon_{c},\eta_{f})$ for given $U,V$) satisfying (from eq.\ \ref{eq:111a})
\begin{equation}
\label{eq:114}
\epsilon_{f}^{*}(\epsilon_{c},\eta_{f})~=~-\epsilon_{f}^{*}(-\epsilon_{c},-\eta_{f})~.
\end{equation}

Since $\rho_{0}(\epsilon)$ is compact, with band edges at $\epsilon = \pm t_{*}$ (eq.\ \ref{eq:11}), it follows 
(eq.\ \ref{eq:112}) that the renormalised levels for any insulating phase necessarily satisfy
\begin{equation}
\label{eq:116}
\Big{|}-\epsilon_{c}+\frac{V^{2}}{\epsilon_{f}^{*}}\Big{|}~ >~ t_{*}~=\tfrac{1}{2}W~~~~~~:~\mathrm{insulators},
\end{equation}
with the system metallic otherwise. 
We will invoke this condition many times in the following.


\subsection{Fermi liquid phases}
\label{subsection:FLphases}

The condition eq.\ \ref{eq:116} for an insulator is general, but does not distinguish between Kondo and Mott insulators. With that in mind we turn to the Fermi liquid regimes of the model -- those adiabatically connected to the non-interacting limit. 
A key result here~\cite{RajaEPJB} is that the total charge is related to the renormalised level by~\cite{fnepsf*=0}
\begin{equation}
\label{eq:117}
\tfrac{1}{2} n_{t}{\pd}~=~\int_{-\infty}^{-\epsilon_{c}+\frac{V^{2}}{\epsilon_{f}^{*}}}d\epsilon ~ \rho_{0}(\epsilon)~+~
\theta(-\epsilon_{f}^{*})
\end{equation}
(with $\theta(x)$ the unit step function). This follows~\cite{RajaEPJB}  from 
eqs.\ \ref{eq:14}-\ref{eq:17}, combined with the central fact that the Luttinger integral~\cite{LW1960,*Luttingerf1960FermiSurf,*Luttingerf1961,LTG2014}
\begin{equation}
\label{eq:118}
I_{L}~:=~\mathrm{Im}\int_{-\infty}^{0}d\w ~G_{f}(\w) \frac{\partial\Sigma_{f}(\w)}{\partial\w}
\end{equation}
vanishes throughout a Fermi liquid phase, $I_{L}=0$, reflecting perturbative continuity to the non-interacting limit.
From the perspective of the effective quantum impurity model, eq.\ \ref{eq:117} amounts to a Friedel sum 
rule.~\cite{Langreth1966,LTG2014}

\begin{figure}
\includegraphics{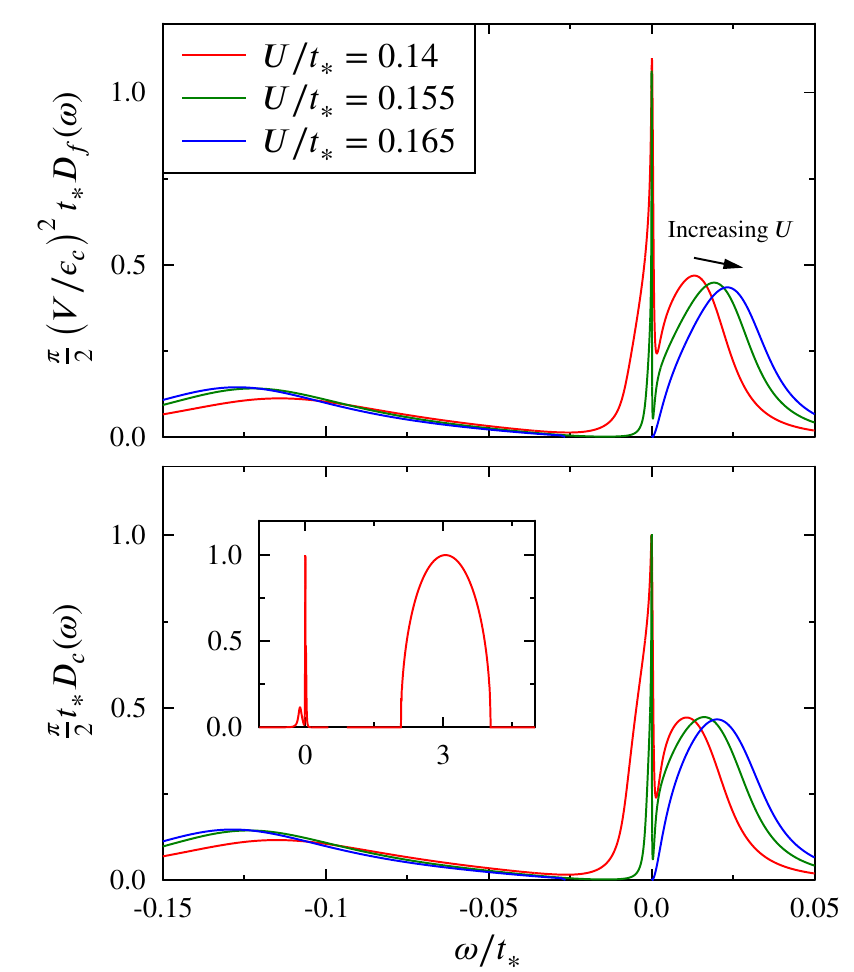}
\caption{\label{fig:fig1} 
Illustrative single-particle spectra for the PAM,  $\tfrac{\pi}{2}t_{*}(V/\epsilon_{c})^{2}D_{f}(\w)$ and
$\tfrac{\pi}{2}t_{*}D_{c}(\w)$, \emph{vs} $\w/t_{*}$. Shown for fixed  $\epsilon_{c}/t_{*}=3$, $V/t_{*}=0.4$ and 
$\eta_{f}=0$, on increasing $U/t_{*}$  towards and through the Mott transition: $U/t_{*} = 0.14$  and $0.155$ 
in the metal, and $0.165$ in the MI (the critical $U_{c}/t_{*} \simeq 0.160$). On approaching the transition 
a low-energy metallic Kondo resonance arises, centred on the Fermi level, here lying  at the lower edge of the upper 
`Hubbard band'. The resonance narrows progressively, vanishing on the spot at the transition, to leave a MI with a fully formed gap. Eq.\ \ref{eq:119} for the metallic phase is obeyed accurately by the NRG results. 
\emph{Lower inset:} $c$-spectrum at higher energies, centred on $\w =\epsilon_{c}$ with band halfwidth $t_{*}$ (there are
no higher energy features in the $f$-spectrum).
}
\end{figure}

From the condition eq.\ \ref{eq:116},  the only non-trivial insulator that can arise from eq.\ \ref{eq:117}
is clearly the $n_{t}=2$ KI (which is perturbatively connected to the $U=0$ hybridization-gap insulator, 
and is thus a Fermi liquid).
For $|-\epsilon_{c}+V^{2}/\epsilon_{f}^{*}|<t_{*}$ by contrast the system is necessarily metallic, with
$D_{\nu}(0) \neq 0$ (eq.\ \ref{eq:112}) and non-integral $n_{t}$. There is thus
no hint of a MI phase here; indeed from eq.\ \ref{eq:117}, $n_{t}=1$ (or $3$) arises only for
$\epsilon_{c}=V^{2}/\epsilon_{f}^{*}$, for which (eq.\ \ref{eq:112})
$D_{c}(0)=\rho_{0}(0)= (V/\epsilon_{c})^{2}D_{f}(0)$, or equivalently 
\begin{equation}
\label{eq:119}
\tfrac{\pi}{2}t_{*}D_{c}(0)~=~1~=~ \tfrac{\pi}{2}t_{*}\left(\frac{V}{\epsilon_{c}}\right)^{2}D_{f}(0)~~~~:~n_{t}=1,3
\end{equation}
-- corresponding to a metal. This reflects the fact that the MI is not adiabatically connected to the non-interacting limit,
whence the Luttinger theorem $I_{L}=0$ does not hold. To handle this non-Fermi liquid insulator requires a two-self-energy description,~\cite{LTG2014,DELMRG2016} considered in sec.\ \ref{section:MottPhase}

Eq.\ \ref{eq:119} does of course capture the approach to the Mott transition from the \emph{metallic},
Fermi liquid side. Here, the $f$-level spectrum $D_{f}(\w)$  contains a low-energy Kondo resonance centred on the Fermi level (reflecting a metallic Kondo effect). The width of the resonance is characterised by a low-energy 
scale $\w_{\mathrm{L}}$ (proportional to the quasiparticle weight
$Z=[1-(\partial\Sigma_{f}^{R}(\w)/\partial\w)_{\w =0}]^{-1}$), which decreases continuously and vanishes as
as the transition is approached from the metal, $n_{t} \rightarrow 1$; with the metallic phase spectra at the Fermi level 
given by eq.\ \ref{eq:119}. At the transition the Kondo resonance then collapses `on the spot', to leave the MI with a fully formed spectral gap. An explicit example is given in fig.\ \ref{fig:fig1}, showing NRG results for both $f$- and $c$-level 
single-particle spectra as the Mott transition is approached and crossed 
(we return to and discuss this figure several times in later sections).

Although eq.\ \ref{eq:117} is confined to Fermi liquid phases, it proves important in understanding the 
behaviour of the PAM phase diagram, to which we now turn.


\subsection{Phase diagram: overview}
\label{subsection: Phasediag}

In considering the phase diagram, our main interest is its dependence on  $\epsilon_{c}$, $U$ and $\eta_{f}$, in particular for the M/MI (Mott) transition (where the hybridization $V$ actually scales out of the problem as shown in 
sec.\ \ref{subsubsection:pbcollapse}). Its qualitative behaviour can in fact be deduced on general grounds, using little more than results already given.

\begin{figure}
\includegraphics{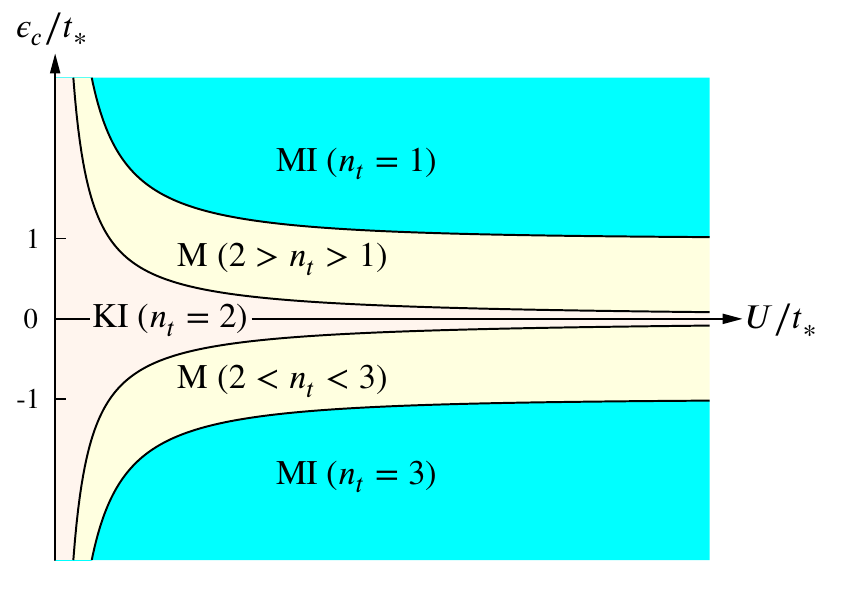}
\caption{\label{fig:fig2} 
Schematic phase boundary in the $(U/t_{*},\epsilon_{c}/t_{*})$-plane, shown for asymmetry $\eta_{f} =0$.
Kondo insulator (KI), metallic (M) and Mott insulator (MI) phases are indicated. The KI/M phase boundary corresponds to a total
charge $n_{t}=n_{f}+n_{c}=2$, and the M/MI boundary to $n_{t}=1$ for $\epsilon_{c} >0$ and $n_{t}=3$ for $\epsilon_{c} <0$.
Note that the MI phase arises only for $|\epsilon_{c}|/t_{*}>1$ (as is generic for $|\eta_{f}|<1$).
For $\eta_{f} =0$ the phase boundaries are precisely symmetric under $\epsilon_{c} \leftrightarrow -\epsilon_{c}$, as indicated.
}
\end{figure}

Fig.\ \ref{fig:fig2} shows the schematic phase boundary in the $(U/t_{*},\epsilon_{c}/t_{*})$-plane, for $f$-level 
asymmetry $\eta_{f} =0$ ($\epsilon_{f}=-U/2$), with the boundaries between the KI/M and M/MI indicated. For $\eta_{f} =0$ 
all phase boundaries are symmetric under $\epsilon_{c} \leftrightarrow -\epsilon_{c}$ (eq.\ \ref{eq:110}), as indicated, so 
only $\epsilon_{c} \geq 0$ need be considered in the following.

The key features of the phase diagram are as follows:\\
\noindent (a) On increasing the interaction $U$ for any fixed $\epsilon_{c}>t_{*}$, or on increasing $\epsilon_{c}$ from $0$ for any fixed $U>0$, three successive phases arise: a KI with $n_{t}=2$, followed by a metal in which $n_{t}$ varies continuously from $2$ to $1$, and finally a MI phase throughout which $n_{t}=1$. 
Each phase occupies a finite fraction of the parameter space, i.e.\ none is `special' in the sense of requiring parametric fine-tuning. No further phases arise here (e.g.\ a metal with $n_{t} <1$, or a band insulator with $n_{t}=0$).~\\
\noindent (b) The phase boundaries decrease monotonically with increasing $U$ (as follows under the natural assumption that 
$n_{t} \equiv n_{t}(\epsilon_{c},U)$ in the M phase decreases monotonically on increasing $\epsilon_{c}$ or $U$).\\
\noindent (c) In the strong coupling limit $U\rightarrow \infty$, $\epsilon_{c}$ for the M/MI phase boundary asymptotically approaches $t_{*} =\tfrac{1}{2}W$, while $\epsilon_{c}$ for the KI/M phase boundary vanishes. \\
\noindent (d) In weak coupling $U\rightarrow 0$ by contrast, $\epsilon_{c}$ for both phase boundaries diverges.

The behaviour above is readily understood as follows:
\begin{enumerate}
\item [{\bf{(i)\rm}}]
Since $\epsilon_{f}=-U/2$, it vanishes in the non-interacting limit (where $\epsilon_{f}^{*} \equiv \epsilon_{f}$). Hence from eq.\ \ref{eq:117},~\cite{fnepsf*=0} $n_{t}=2$ for \emph{all} $\epsilon_{c}$, i.e.\ the system is a KI 
along the entire $\epsilon_{c}$-axis for $U=0$ (which gives the origin of (d) above). The system is likewise a KI along the entire $U$-axis for $\epsilon_{c}=0$, as follows from eq.\ \ref{eq:117} on noting from eq.\ \ref{eq:114} that 
$\epsilon_{f}^{*}(\epsilon_{c}=0, \eta_{f} =0) =0$. 
Neither phase boundary can therefore intersect the axes; a KI thus persists for a finite range of $U$, $\epsilon_{c}$ away from them. 
\item [{\bf{(ii)\rm}}]
Now consider $U\rightarrow \infty$, remembering that $\epsilon_{f}=-U/2$ and $\epsilon_{f}+U=+U/2$.
Here the $f$-levels are asymptotically singly occupied only (with doubly-occupied or empty $f$-levels precluded).
Hence $n_{f} =1$, and $n_{c}\equiv n_{c}^{0}$ is equivalently that for the free conduction band 
($V$ is obviously asymptotically inoperative as $U\rightarrow \infty$); i.e.\ 
\begin{subequations}
\label{eq:120}
\begin{align}
n_{t}~=&~1 ~+~ n_{c}^{0}~~~~~~:~U \rightarrow \infty
\\
\tfrac{1}{2}n_{c}^{0}~=&~ \int^{-\epsilon_{c}}_{-\infty}d\epsilon~\rho_{0}(\epsilon)
\end{align}
\end{subequations}
From this, $n_{c}^{0}=1$ and hence $n_{t}=2$ (a KI) arises only for $\epsilon_{c}=0$.
By contrast, $n_{c}^{0}=0$ and hence $n_{t}=1$ (a MI) arises for all $\epsilon_{c} >t_{*}$ (the band edges of 
$\rho_{0}(\epsilon)$ occur at $\pm t_{*}$). These give the $U \rightarrow \infty$ asymptotes for the KI/M and M/MI 
boundaries in  (c) above (as shown in fig.\ \ref{fig:fig2}).
\item [{\bf{(iii)\rm}}]
Consider now any given $U >0$, as $\epsilon_{c} \rightarrow \infty$. Here the conduction band is 
`projected out', so the $c$-level charge $n_{c}$ obviously vanishes and the hybridization $V\equiv 0$ is inoperative. 
The correlated $f$-levels are thus free and singly-occupied only, $n_{f}=1$; whence  $n_{t}=1$, corresponding to a MI. 
This is why no further phases arise in fig.\ \ref{fig:fig2}, as noted in (a) above.
\end{enumerate}

One further observation bears note here. As argued above, the critical $\epsilon_{c}(U)$ for the Mott transition exceeds 
$t_{*}$ for all finite $U$, and indeed diverges as $U\rightarrow 0$. In otherwords the metal persists to values of 
$\epsilon_{c}$ far in excess of the Fermi level, where one might naively expect the conduction band to empty and the system to be insulating; and the corresponding critical $U/t_{*}$ for the transition asymptotically vanishes, i.e.\ 
appears in `weak coupling' -- somewhat counterintuitively for a transition regarded as an archetype of strong correlations. 
This is striking, suggests that the transition is a fairly subtle affair, and invites an explanation. We provide it in 
secs.\ \ref{section:Hubbardmapping} \emph{ff}.


\subsubsection{Renormalised levels}
\label{subsubsection:renormlevels1}

The phase boundaries correspond of course to specific values of $n_{t}$. Eq.\ \ref{eq:117} thus generically determines the renormalised levels $\epsilon_{f}^{*}$ along them; or more precisely, in the case of the M/MI boundary, as the transition is approached from the M side (remembering that eq.\ \ref{eq:117} holds strictly for FL phases).
 Note from eq.\ \ref{eq:117} that $n_{t}\leq 2$ for $\epsilon_{f}^{*} \geq 0$, whence $\epsilon_{f}^{*} \geq 0$
for $\epsilon_{c} \geq 0$. For the M/MI border, where $n_{t} \rightarrow 1+$, 
eq.\ \ref{eq:117} requires $-\epsilon_{c}+V^{2}/\epsilon_{f}^{*}=0$, i.e.\
\begin{equation}
\label{eq:121}
\epsilon_{f}^{*}~=~ \frac{V^{2}}{\epsilon_{c}}~~~~~~:~ \mathrm{M/MI~boundary,~}\epsilon_{c} \geq t_{*}.
\end{equation}
For the KI/M border, eq.\ \ref{eq:117} gives $-\epsilon_{c}+V^{2}/\epsilon_{f}^{*}=t_{*}$, i.e.\
\begin{equation}
\label{eq:122}
\epsilon_{f}^{*}~=~ \frac{V^{2}}{\epsilon_{c}+t_{*}}~~~~~~:~ \mathrm{KI/M~boundary,~}\epsilon_{c} \geq 0.
\end{equation}
Referring then to fig.\ \ref{fig:fig2}, on increasing the interaction for any fixed $\epsilon_{c}>t_{*}$ the renormalised level thus progressively increases from $(\epsilon_{f} \equiv)$ $\epsilon_{f}^{*}=0$ for $U=0$, to  
$\epsilon_{f}^{*}=V^{2}/(\epsilon_{c}+t_{*})$ at the KI/M boundary, to $\epsilon_{f}^{*}=V^{2}/\epsilon_{c}$ as the M/MI transition is approached from the metal.

Results for $\epsilon_{c}\leq 0$ naturally follow in a directly analogous way, giving
$\epsilon_{f}^{*}=-V^{2}/(|\epsilon_{c}|+t_{*})$ for the KI/M boundary, and $\epsilon_{f}^{*}=-V^{2}/|\epsilon_{c}|$ for the 
M/MI border. For $\eta_{f} =0$ in particular (fig.\ \ref{fig:fig2}), these reflect the symmetry 
$\epsilon_{f}^{*}(\epsilon_{c},0)=-\epsilon_{f}^{*}(-\epsilon_{c},0)$ for given $U$ (eq.\ \ref{eq:114}).~\cite{KIMboundary}

We return to these results in sec.\ \ref{subsection:renormnormal} (fig.\ \ref{fig:fig10}), after 
$\epsilon_{f}^{*}$ in the MI phase has been considered.


\subsection{Phase diagram: take two}
\label{subsection: Phasediagtake2}

While we have thus far focused on $\eta_{f} =0$, the same essential characteristics above arise also for 
$|\eta_{f}|<1$ (although for $\eta_{f} \neq 0$ the phase boundaries are obviously no longer strictly symmetric under 
$\epsilon_{c}\leftrightarrow -\epsilon_{c}$ at fixed $U$). Here, $\epsilon_{f} =-\tfrac{1}{2}U(1-\eta_{f}) <0$ and 
$\epsilon_{f}+U =\tfrac{1}{2}U(1+\eta_{f})>0$ for all $U>0$, so the arguments given in {\bf{(ii)\rm}} and {\bf{(iii)\rm}} above again go through. Likewise, by the same reasoning as in {\bf{(i)\rm}} above, the system is a KI along the entire 
$\epsilon_{c}$-axis for $U=0$ (so for $\epsilon_{c}=0$ in particular, $n_{t}=2$ (KI) for both the $U=0$ and 
$U\rightarrow \infty$ asymptotes).

There is a further notable feature in fig.\ \ref{fig:fig2}, occurring generally for $|\eta_{f}|<1$: the Mott transition to the 
$n_{t}=1$ insulator arising for $\epsilon_{c}>0$ occurs \emph{solely} from the electron-doped side, i.e.\ $n_{t}=1+\delta$ with $\delta \rightarrow 0+$ [and correspondingly for $\epsilon_{c}<0$, the transition to the $n_{t}=3$ MI
occurs only from the hole-doped side, $n_{t}=3+\delta$ with $\delta \rightarrow 0-$]. This reflects the fact, 
see {\bf{(iii)\rm}} above, that the $n_{t}=1$ MI persists as $\epsilon_{c} \rightarrow \infty$ for any $U>0$. 

\begin{figure}
\includegraphics{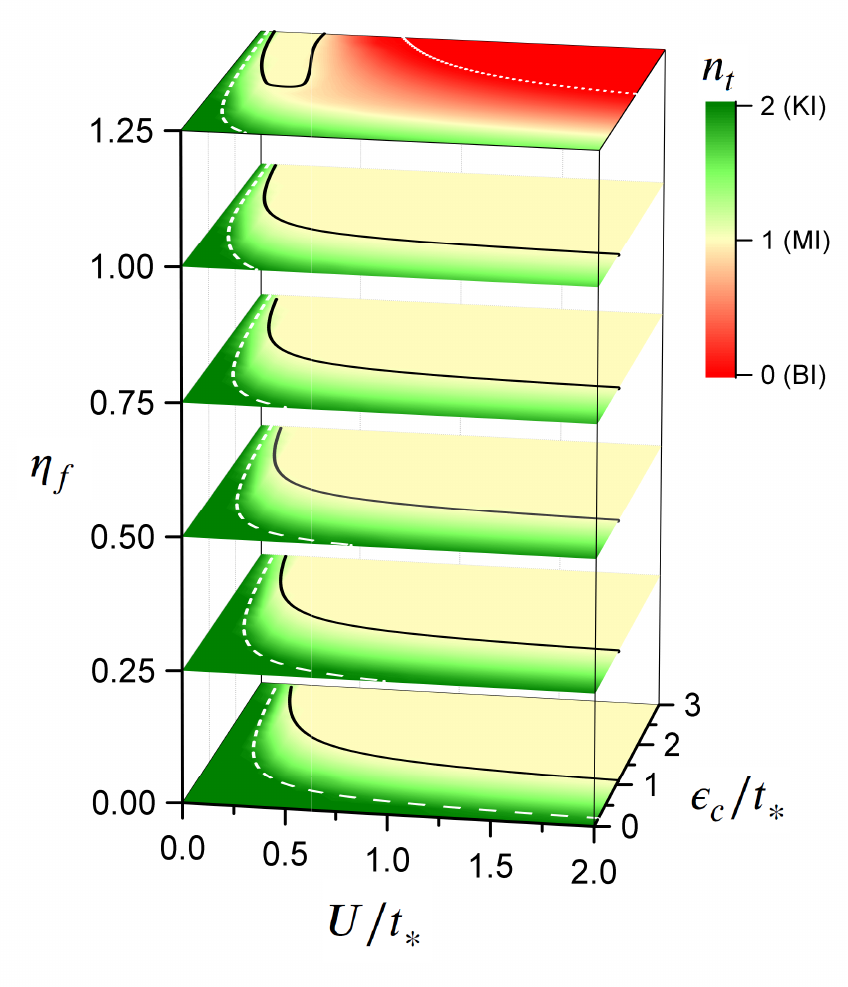}
\caption{\label{fig:fig3} 
NRG-determined phase boundaries, shown in the ($U/t_{*},\epsilon_{c}/t_{*}$)-plane for different values of the $f$-level asymmetry $\eta_{f} =0, \tfrac{1}{4}, \tfrac{1}{2}, \tfrac{3}{4}, 1, \tfrac{5}{4}$ (from bottom to top). $V/t_{*}=0.4$ is fixed, and only $\epsilon_{c}/t_{*} \geq 0$ is shown for clarity. The charge $n_{t}$ is colour coded as indicated.
The transition to the $n_{t}=1$ Mott insulator is shown as a solid line, the KI/M transition as dashed,
and the metal to $n_{t}=0$ band insulator transition is the dotted line (as relevant to the $\eta_{f}=\tfrac{5}{4}$ case).
}
\end{figure}

That situation changes for asymmetries $|\eta_{f}| >1$, the $f$-level mixed-valence regime.
For $\eta_{f}>1$, $\epsilon_{f} = -\tfrac{1}{2}U(1-\eta_{f})$ and $\epsilon_{f}+U$ both exceed $0$ (the Fermi level).
Hence, following the arguments in {\bf{(iii)\rm}} above, as $\epsilon_{c} \rightarrow \infty$ for any given $U>0$, the 
$f$-levels become free but are now \emph{un}occupied, with $n_{f}=0$ ($=n_{c}$): the system is thus asymptotically an 
$n_{t}=0$ band insulator [or an $n_{t}=4$ band insulator from the obvious argument for $\epsilon_{c} \rightarrow -\infty$].
For this reason the $n_{t}=1$ MI sector of the phase diagram is then surrounded on the higher-$\epsilon_{c}$ side by a metal
with $n_{t}<1$ (which for larger $\epsilon_{c}$ undergoes a transition to the $n_{t}=0$ band insulator); and on the 
lower-$\epsilon_{c}$ side by a metal with $n_{t}>1$. In this case the Mott transition can thus be approached from either the electron-doped or the hole-doped sides, $n_{t}\rightarrow 1\pm$ respectively; although we add that, in contrast to the Mott transition in the one-band Hubbard model, the two are not in any way related by a ph-transformation. \\

\comment{
\begin{figure}
\includegraphics{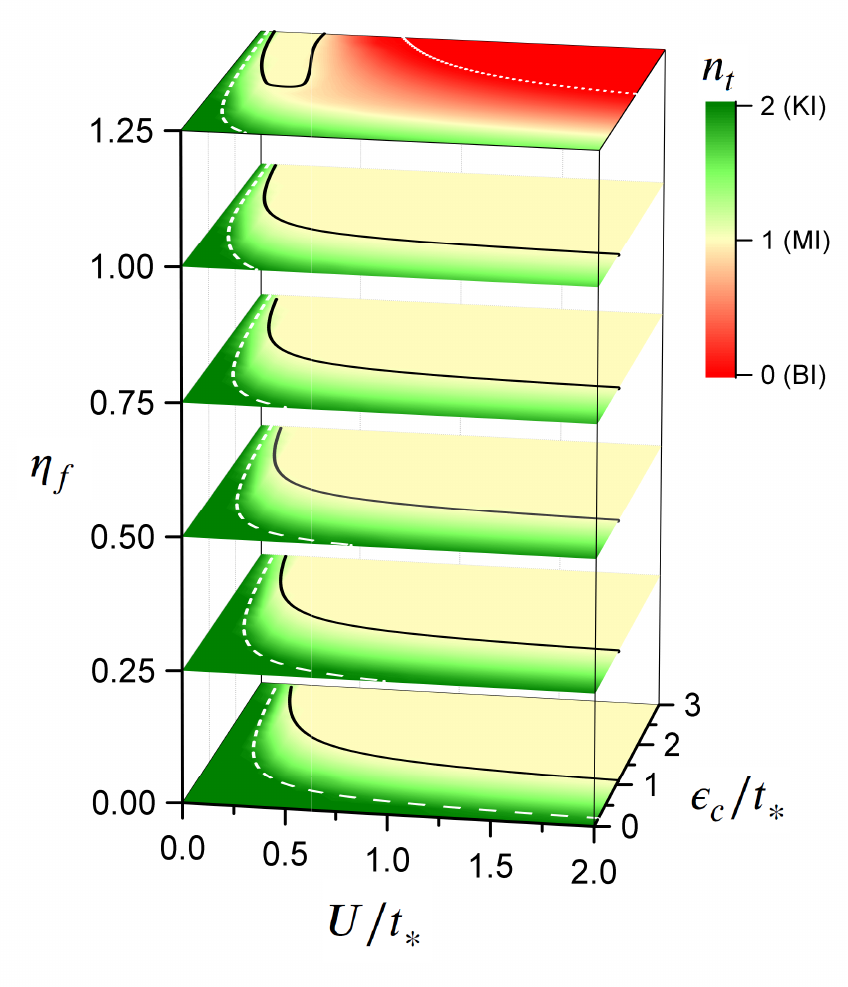}
\caption{\label{fig:fig3} 
NRG-determined phase boundaries, shown in the ($U/t_{*},\epsilon_{c}/t_{*}$)-plane for different values of the $f$-level asymmetry $\eta_{f} =0, \tfrac{1}{4}, \tfrac{1}{2}, \tfrac{3}{4}, 1, \tfrac{5}{4}$ (from bottom to top). $V/t_{*}=0.4$ is fixed, and only $\epsilon_{c}/t_{*} \geq 0$ is shown for clarity. The charge $n_{t}$ is colour coded as indicated.
The transition to the $n_{t}=1$ Mott insulator is shown as a solid line, the KI/M transition as dashed,
and the metal to $n_{t}=0$ band insulator transition is the dotted line (as relevant to the $\eta_{f}=\tfrac{5}{4}$ case).
}
\end{figure}
}

That the characteristics deduced in the sections above are indeed found in NRG calculations is seen clearly in 
fig.\ \ref{fig:fig3}, where resultant phase diagrams are shown in the ($U/t_{*},\epsilon_{c}/t_{*}$)-plane for six representative values of the $f$-level asymmetry $\eta_{f}$. NRG  phase boundaries between metallic and insulating phases are 
determined~\cite{DELMRG2016}  by approaching a transition from the metal, and locating the points where the total charge 
$n_{t} \rightarrow 1$, $2$ or $0$ for the metal to MI, KI and band insulator phases respectively.
Note that the $\eta_{f} =5/4$ case in particular shows all possible phases -- metals with $n_{t}\gtrless1$,
MI, KI and band insulators.

\begin{figure}
\includegraphics{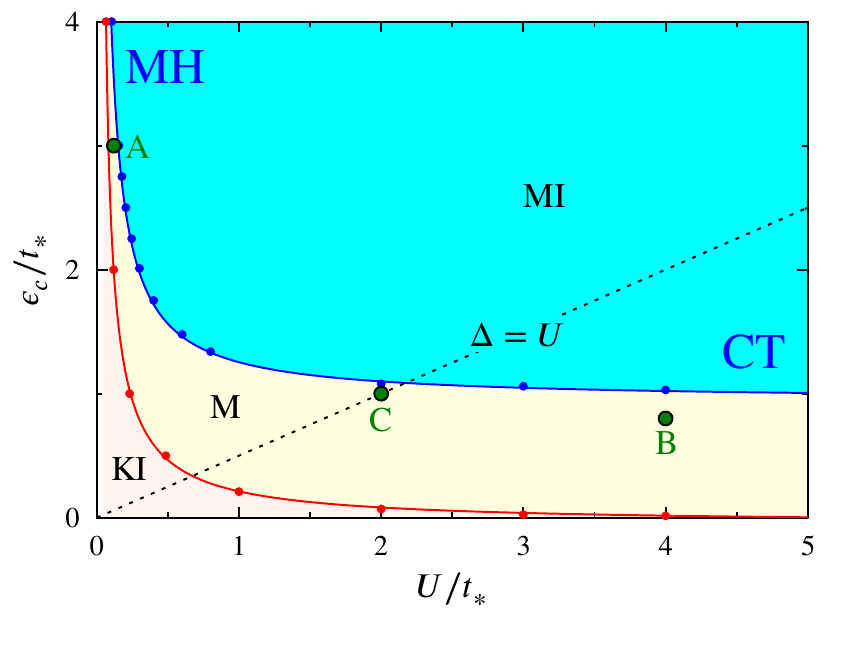}
\caption{\label{fig:fig4} 
NRG phase diagram for $\eta_{f}=0$ (with $V/t_{*}=0.4$), showing the crossover line with charge transfer energy
$\Delta =U$  (i.e.\ $\epsilon_{c} = \tfrac{1}{2}U$) separating Mott-Hubbard and charge-transfer insulators; 
amounting to a particular example of a Zaanen-Sawatzky-Allen diagram.~\cite{Zaanen-S-A1985} Discussion in text.
Points A, B and C indicate metals with $\Delta \gg U$, $\Delta \ll U$ and $\Delta \simeq U$ respectively;
considered as representative examples in secs.\ \ref{section:spdynamics} (single-particle dynamics for point A are shown in figs.\ \ref{fig:fig1} and \ref{fig:fig7}, and for point B in fig.\ \ref{fig:fig8}).
}
\end{figure}

The results above also provide concrete examples of Zaanen-Sawatzky-Allen (ZSA) diagrams,~\cite{Zaanen-S-A1985} 
with Mott insulators classified as either of `Mott-Hubbard' (MH) or `charge-transfer' (CT) type; according respectively to whether $U<\Delta$ or $U>\Delta$, with
$\Delta =|\epsilon_{c}-\epsilon_{f}| = |\epsilon_{c}+\tfrac{1}{2}U(1-\eta_{f})|$ the charge transfer energy.
Fig.\ \ref{fig:fig4} shows an NRG phase diagram for $\eta_{f}=0$, including the (crossover) line $\Delta =U$ separating MH and CT insulators, \emph{viz} $\epsilon_{c, \mathrm{cr}} = \epsilon_{f}+U \equiv \tfrac{1}{2}U$.
For transition metal (TM) materials, oxides or halides of the lighter TMs tend to be MH insulators, while 
those of the heavier TMs are more typically of CT type.~\cite{Zaanen-S-A1985} 
Metallic TM compounds with  $\Delta \gg U$ are `$d$-band metals' (again usually for the lighter TMs), characterised by 
narrow bands and heavy electrons/holes;~\cite{Zaanen-S-A1985} while those with $\Delta \ll U$ are  `$p$-band metals', 
with light carriers and broad bands.

In the following sections, within the periodic Anderson model under study, we aim to shed some light on how and why these and concomitant characteristics arise.

\section{Mapping to a Hubbard model}
\label{section:Hubbardmapping}

It has hitherto been thought that the low-energy properties of the metallic phase close to the Mott transition 
cannot be interpreted generally in terms of a single-band Hubbard model.~\cite{SordiPRL2007,*SordiPRB2009,AmaricciPRB2012} 
Here, however, we show that the effective low-energy model which describes the PAM in the vicinity of the Mott transition, is 
in fact a one-band Hubbard model, with long-ranged hopping between the sites; specifically hoppings that connect 
all pairs of sites rather than simply nearest neighbours, and which decay exponentially with the topological distance between
them. The plausibility of generating an effective one-band model is intuitive, since  on integrating out virtual excitations to
high-lying $c$-levels one expects to generate effective hoppings between $f$-levels.
There are in fact several ways to obtain these results, and here we do so by direct analysis of the underlying propagators.
We also focus in the following on the Mott transition for $n_{t}=1$, which 
(secs.\ \ref{subsection: Phasediag},\ref{subsection: Phasediagtake2}) requires $\epsilon_{c}>t_{*}$ (the $n_{t}=3$ case obviously follows in the same way by considering $\epsilon_{c} <-t_{*}$, and adds nothing new).


\subsection{One-band Feenberg self-energy}
\label{subsection:FSEoneband}

To this end we first point out a general condition that must be satisfied by the Feenberg 
self-energy~\cite{Feenberg,*Economou} for a one-band model. Within DMFT, the local propagator for any one-band 
model (here denoted generically by $G(\w)$) is of form
\begin{equation}
\label{eq:123}
G(\w)~=~ \int^{\infty}_{-\infty}d\epsilon~ \frac{\tilde{\rho}(\epsilon)}{\w^{+}-\epsilon_{0} -\Sigma(\w) -\epsilon}
\end{equation}
with $\epsilon_{0}$ the site-energy, $\Sigma(\w)$ the (local) self-energy, and $\tilde{\rho}(\epsilon)$ the non-interacting DoS for the one-band model. The corresponding Feenberg self-energy $S(\w)$ is (by its definition) related to $G(\w)$ by
\begin{equation}
\label{eq:124}
G(\w)~=~ \left[
\w^{+}-\epsilon_{0} -\Sigma(\w) - S(\w)
\right]^{-1}
\end{equation}
(e.g.\ $S(\w) = \tfrac{1}{4}\tilde{t}^{2}_{*}G(\w)$ for a NN Hubbard model on a Bethe lattice, 
with hopping $\tilde{t}_{*}$). Eqs.\ \ref{eq:123},\ref{eq:124} are general; what distinguishes different one-band models 
is of course the associated DoS $\tilde{\rho}(\epsilon)$. From eqs.\ \ref{eq:123},\ref{eq:124} it follows that
\begin{equation}
\label{eq:125}
G(\w)~=~ \int^{\infty}_{-\infty}d\epsilon~ \frac{\tilde{\rho}(\epsilon)}{\frac{1}{G(\w)} +S(\w) -\epsilon}.
\end{equation}
Hence $S \equiv S(G(\w))$; i.e.\ the $\w$-dependence of the one-band Feenberg self-energy arises solely from its
dependence on $G(\w)$, with no explicit $\w$-dependence otherwise. This necessary condition for a one-band model is
exploited below.


\subsection{Effective one-band model}
\label{subsection:lowenergyHM}

Since the Mott transition occurs generally only for $\epsilon_{c}>t_{*}$  (secs.\ 
\ref{subsection: Phasediag},\ref{subsection: Phasediagtake2}), it is this regime on which we 
focus in the following. Consider first the propagator $\gc$ for the free ($V=0$) PAM $c$-band, given 
(eq.\ \ref{eq:18a}) by
\begin{equation}
\label{eq:126}
\gc ~=~\left[\w^{+}-\epsilon_{c} - \tfrac{1}{4}t^{2}_{*}\gc\right]^{-1}~.
\end{equation}
The corresponding spectrum $d_{0}^{c}(\w) =\rho_{0}(\w -\epsilon_{c})$ is non-zero for $|\w-\epsilon_{c}|<t_{*}$; 
so since $\epsilon_{c} > t_{*}$, $d_{0}^{c}(\w =0)$ in particular vanishes and $g_{c}(0)$ is pure real,
\begin{subequations}
\label{eq:127}
\begin{align}
\tfrac{1}{2}t_{*}g_{c}(0) ~=~ &
-\frac{\epsilon_{c}}{t_{*}} +\sqrt{\left(\frac{\epsilon_{c}}{t_{*}}\right)^{2}-1}
\label{eq:127a}
\\
\overset{\epsilon_{c}/t_{*} \gg 1}{\sim} &-\frac{t_{*}}{2\epsilon_{c}}.
\label{eq:127b}
\end{align}
\end{subequations}

We seek $\Gf$ in the vicinity of the Fermi level, and to that end 
consider the $f$-level Feenberg self-energy $S_{f}(\w)=V^{2}[\w^{+}-\epsilon_{c} -\tfrac{1}{4}t_{*}^{2}\Gc]^{-1}$ 
(eq.\ \ref{eq:18}). Separate 
\begin{equation}
\label{eq:128}
\Gc~=~\gc ~+~ \delta\Gc
\end{equation}
where $\gc$ is pure real for $\w \approx 0$, so the spectral weight in $\Gc$  resides
entirely in $\delta\Gc$. Hence, using eq.\ \ref{eq:126}, 
\begin{equation}
\label{eq:129}
S_{f}(\w)~=~ \frac{V^{2}\gc}{1-\tfrac{1}{4}t_{*}^{2}\gc\delta\Gc}.
\end{equation}
We thus consider $\delta\Gc$, and first define the auxilliary functions
\begin{equation}
\begin{split}
\tilde{g}_{c}(\w)=&\left[
\w^{+}-\epsilon_{c} - \tfrac{1}{4}t^{2}_{*}\Gc
\right]^{-1}
\\
g_{f}(\w)=&\left[
\w^{+}-\epsilon_{f} - \Sigf
\right]^{-1}.
\end{split}
\nonumber
\end{equation}
Then eqs.\ \ref{eq:18} give the identity
\begin{equation}
\label{eq:130}
\Gc ~=~ \gctil ~+~[V\gctil]^{2}\Gf~.
\end{equation}
Separating $\Gc$ as in eq.\ \ref{eq:128}, and noting that $\gctil = \gc/[1-\tfrac{1}{4}t^{2}_{*}\gc\delta\Gc]$,  
eq.\ \ref{eq:130} gives a simple cubic equation for $t_{*}\delta\Gc$,
\begin{equation}
\label{eq:131}
\begin{split}
&\tfrac{1}{16}(t_{*}g_{c})^{2}[t_{*}\delta\Gc]^{3}+\tfrac{1}{16}t_{*}g_{c}([t_{*}g_{c}]^{2}-8)~[t_{*}\delta\Gc]^{2}
\\
& + (1-\tfrac{1}{4}[t_{*}g_{c}]^{2})~t_{*}\delta\Gc ~=~ (Vg_{c} )^{2} ~t_{*}\Gf
\end{split}
\end{equation}
where $g_{c} \equiv \gc$. In the trivial uncoupled limit of $V=0$, the correct physical root is 
of course  $\delta\Gc =0$.

Eq.\ \ref{eq:131} shows that $\delta\Gc$ -- and hence $S_{f}(\w)$ (eq.\ \ref{eq:129}) -- is in general a function of $\Gf$ \emph{and} $\gc$ (which is an explicit function of $\w$). By the argument given in sec.\ \ref{subsection:FSEoneband}, $\Gf$ does not therefore generically reduce to an effective one-band model, just as one 
expects. There are however two important exceptions, as now considered.

\subsubsection{$\epsilon_{c} \gg t_{*}$}
\label{subsubsection:bigepsilonc}

 For $\epsilon_{c} \gg t_{*}$, the free conduction band (centred on $\w =\epsilon_{c}$) lies so far above 
the Fermi level ($\w =0$) that for all $\w \ll \epsilon_{c} -t_{*}$ -- and in particular for energies around the Fermi 
level -- $\gc$ is independent of $\w$ and given by $g_{c}(0) \sim -1/\epsilon_{c}$ (eq.\ \ref{eq:127b}). In this case 
eq.\ \ref{eq:131} gives to leading order
\begin{equation}
\label{eq:132}
\delta \Gc = \left(\frac{V}{\epsilon_{c}}\right)^{2}~\Gf
\end{equation}
(where only the final term on the left side of eq.\ \ref{eq:131} is relevant); showing that for all 
$\w \ll \epsilon_{c}-t_{*}$, the spectral density of $G_{c}(\w)$ is entirely controlled by that of $G_{f}(\w)$.
Eq.\ \ref{eq:129} then yields asymptotically
\begin{equation}
S_{f}(\w) 
~= -\frac{V^{2}}{\epsilon_{c}} +\frac{1}{4}\left(t_{*}\frac{V^{2}}{\epsilon_{c}^{2}}\right)^{2}\Gf,
\nonumber
\end{equation}
the $\w$-dependence of which is encoded solely in that of $\Gf$. The effective low-energy model is thus one-band, with
eq.\ \ref{eq:18}c reducing to
\begin{equation}
\label{eq:133}
\Gf=\left[
\w^{+} - (\epsilon_{f}-\frac{V^{2}}{\epsilon_{c}}) -\Sigma_{f}(\w) - \tfrac{1}{4}t_{*\mathrm{eff}}^{2}\Gf
\right]^{-1}
\end{equation}
where
\begin{equation}
\label{eq:134}
t_{*\mathrm{eff}}~=~t_{*}\left(
\frac{V}{\epsilon_{c}}
\right)^{2}.
\end{equation}
This corresponds precisely to an effective NN one-band Hubbard model for the $f$-levels alone, with 
a site-energy $\epsilon_{0} = \epsilon_{f}-V^{2}/\epsilon_{c}$ (naturally renormalised from the `bare' $\epsilon_{f}$ 
due to an induced level-shift from the $c$-band), and an effective NN hopping $\teff$. Since the shortest connected 
path between two NN $f$-levels is a 3-step path -- proceeding by virtual hopping to and between the intervening pair of 
$c$-levels -- the third-order perturbative form eq.\ \ref{eq:134} for $\teff$ is physically quite obvious.

\subsubsection{General case: vicinity of the Mott transition}
\label{subsubsection:closetomt}

 The second case where the effective low-energy model for the PAM is of one-band form, is 
quite generally in the vicinity of the Mott transition for \emph{any} $\epsilon_{c}>t_{*}$,
as now shown.  Here the $f$-spectrum in the metallic phase contains a Kondo-like resonance, characterised by 
the low-energy scale $\w_{\mathrm{L}}$ which vanishes at the transition. The scaling regime 
of the transition corresponds to considering any finite $\tilde{\w} =\w/\w_{L}$ in the formal limit $\w_{L} \rightarrow 0$; and as such encompasses the \emph{entire} energy-dependence of the underlying Kondo resonance. In this regime, the propagators $G_{\nu}$ and self-energy $\Sigma_{f}$ are functions solely of $\tilde{\w}$, although we continue to denote them by $G_{\nu}(\w)$ and $\Sigma_{f}(\w)$. By contrast, `bare' factors of $\w \equiv \w_{L}\tilde{\w}$ can of course be neglected; 
e.g.\ eq.\ \ref{eq:18b},\ref{eq:18c} becomes
\begin{subequations}
\label{eq:136}
\begin{align}
G_{f}(\w)=& \left[
i0^{+}-\epsilon_{f} -\Sigma_{f}(\w) -\frac{V^{2}}{i0^{+}-\epsilon_{c} -\tfrac{1}{4}t^{2}_{*}G_{c}(\w)}
\right]^{-1}
\label{eq:136a}
\\
=& \left[
i0^{+}-\epsilon_{f} -\Sigma_{f}(\w) -S_{f}(\w)
\right]^{-1} ,
\label{eq:136b}
\end{align}
\end{subequations}
and likewise $g_{c}(\w) \equiv g_{c}(0)$. Separating $G_{c}(\w) \equiv g_{c}(0) + \delta G_{c}(\w)$, and
running through the argument given in eqs.\ \ref{eq:128}\emph{ff}, leads again to the cubic
eq.\ \ref{eq:131} for $t_{*}\delta G_{c}(\w)$, but now with $g_{c} \equiv g_{c}(0)$;
such that $\delta G_{c}(\w) \equiv \delta G_{c}(G_{f}(\w))$ is a function of $G_{f}(\w)$ alone.
Hence the $\wtil$-dependence of $\delta G_{c}(\w)$, and in consequence of
\begin{equation}
\label{eq:137}
S_{f}(\w) ~\equiv ~ \frac{V^{2}g_{c}(0)}{1-\tfrac{1}{4}t_{*}^{2}g_{c}(0)\delta G_{c}(G_{f}(\w))}~,
\end{equation}
arises solely from that of $G_{f}(\w)$. By the argument given in sec.\ \ref{subsection:FSEoneband}, the effective low-energy model in the vicinity of the Mott transition is thus generally of one-band form. As in eqs.\ \ref{eq:123},\ref{eq:125}, 
to determine what that effective model is requires us to ascertain the corresponding DoS $\tilde{\rho}(\epsilon)$. This we now consider.


\subsection{Long-ranged Hubbard model}
\label{subsection:longrangeHM}

 In the $\wtil$-scaling regime of the Mott transition discussed above, the $f$-propagator is given by
(see eqs.\ \ref{eq:15},\ref{eq:17a})
\begin{equation}
\label{eq:138}
\Gf ~=~ \int^{\infty}_{-\infty}d\epsilon ~\frac{\rho_{0}(\epsilon)}{i0^{+} -\epsilon_{f}
-\Sigf -\frac{V^{2}}{i0^{+}-\epsilon_{c}-\epsilon}}
\end{equation}
(with $\rho_{0}(\epsilon)$ from eq.\ \ref{eq:12}); which again we emphasise encompasses the 
$\wtil$-scaling regime of the entire Kondo resonance. Now define
\begin{subequations}
\label{eq:139}
\begin{align}
\teff ~=&~ t_{*} [Vg_{c}(0)]^{2} ~~~( =\tfrac{4V^{2}}{t_{*}}y^{2})
\label{eq:139a}
\\
y ~=&~ \tfrac{1}{2}t_{*}g_{c}(0),
\label{eq:139b}
\end{align}
\end{subequations}
the physical significance of which will become clear below; where $y$ is a function solely of $\epsilon_{c}/t_{*}$, 
from eq.\ \ref{eq:127a}. Since $g_{c}(0)$ satisfies eq.\ \ref{eq:126}, it follows that
$(i0^{+}-\epsilon_{c})g_{c}(0)=1+[\tfrac{1}{2}t_{*}g_{c}(0)]^{2}$, whence
$V^{2}[i0^{+} -\epsilon_{c}-\epsilon]^{-1} = \tfrac{\teff}{2y} [1+y^{2} - 2y\tfrac{\epsilon}{t_{*}}]^{-1}$.
Eq.\ \ref{eq:138} thus becomes
\begin{equation}
\label{eq:140}
\begin{split}
& \Gf  ~=~ 
\\
&\int^{\infty}_{-\infty}d\epsilon \frac{\rho_{0}(\epsilon)}{i0^{+} -(\epsilon_{f}
+\frac{\teff}{2y})
-\Sigf - \frac{\teff}{2y}
\left(\frac{1}{(1+y^{2})-2y\frac{\epsilon}{t_{*}}}-1\right)
}
\end{split}
\end{equation}
Now simply change integration variables from $\epsilon$ to $\epsilon^{\prime}$, defined by
\begin{equation}
\epsilon^{\prime}~=~ \frac{\teff}{2y}\left(
\frac{1}{(1+y^{2})-2y\frac{\epsilon}{t_{*}}}-1
\right),
\nonumber
\end{equation}
and with  $\epsilon \equiv \epsilon(\epsilon^{\prime})$  given on inversion by
\begin{equation}
\label{eq:141}
\begin{split}
\epsilon~=&~ \frac{t_{*}}{2y}\left(
[1+y^{2}] -\frac{1}{1+\frac{2y}{\teff}\epsilon^{\prime}}
\right)
\\
=&~ t_{*} \lambda (\epsilon^{\prime})
\end{split}
\end{equation}
and $\lambda(\epsilon^{\prime})$ thus defined. 
With this 
(using $\teff/2y = V^{2}g_{c}(0)$ from eq.\ \ref{eq:139}), eq.\ \ref{eq:140} reduces to
\begin{equation}
\label{eq:142}
\Gf = \int^{\infty}_{-\infty}d\epsilon ~\frac{\tilde{\rho}(\epsilon)}
{i0^{+}- \left(\epsilon_{f} +V^{2}g_{c}(0)\right)-\Sigf
-\epsilon
}
\end{equation}
where
\begin{equation}
\label{eq:143}
\tilde{\rho}(\epsilon)~=~ \frac{1}{\Big(1+\frac{2y}{\teff}\epsilon\Big)^{2}}~\frac{2}{\pi\teff}
\sqrt{1 -\lambda^{2}(\epsilon)}~.
\end{equation}
Eq.\ \ref{eq:143} is precisely the density of states for a one-band tight-binding model on a 
($Z_{c}\rightarrow \infty$) Bethe lattice with exponentially decreasing hopping amplitudes, as considered 
in Ref.\ [\onlinecite{Eckstein2005}]; such that the $n^{\mathrm{th}}$ NN hopping amplitude, $\teff^{(n)}$, is
\begin{equation}
\label{eq:144}
\teff^{(n)}~=~
\teff ~y^{n-1}~=~
t_{*} \left(\frac{2V}{t_{*}}\right)^{2} y^{n+1}
\end{equation}\\
with a nearest neighbour ($n=1$) hopping amplitude of $\teff = t_{*}[Vg_{c}(0)]^{2}$ (as in eq.\ \ref{eq:139a}).

Eq.\ \ref{eq:142} is thus the local propagator for a one-band Hubbard model with this long-ranged hopping; and an effective site-energy $\epsilon_{0}$ (eq.\ \ref{eq:123}) given  on comparison of eqs.\ \ref{eq:123},\ref{eq:142} by
\begin{subequations}
\label{eq:145}
\begin{align}
\epsilon_{0}~=&~\epsilon_{f} ~+~V^{2}g_{c}(0)
\label{eq:145a}
\\
=&~\epsilon_{f}~+~\frac{2V^{2}}{t_{*}}y ~=~\epsilon_{f}~+~\frac{\teff}{2y}
\label{eq:145b}
\end{align}
\end{subequations}
(with  $\tfrac{1}{2}t_{*}g_{c}(0)=y$ pure real for all $\epsilon_{c} >t_{*}$, eq.\ \ref{eq:127}).
Physically, $\epsilon_{0} \neq \epsilon_{f}$ reflects the renormalization of the site-energy from the `bare' 
$\epsilon_{f}$  due to level-repulsion from the $c$-band. Note that the general case considered here reduces to the results 
of sec.\ \ref{subsubsection:bigepsilonc}, in the regime $\epsilon_{c}/t_{*} \gg 1$ where the low-energy Hubbard model
becomes in effect pure NN, with  $\epsilon_{0} =\epsilon_{f} -V^{2}/\epsilon_{c}$.

Now consider briefly the non-interacting effective Hubbard model's spectrum $\tilde{\rho}(\epsilon)$,
eq.\ \ref{eq:143}, which is clearly ph-asymmetric in general 
($\tilde{\rho}(\epsilon)\neq \tilde{\rho}(-\epsilon)$).~\cite{Eckstein2005} Its band edges in particular occur at 
$\epsilon = W_{+}$ and $\epsilon =-W_{-}$, with $W{+} \neq W_{-}$ in general. These are given simply by
\begin{equation}
\label{eq:146}
W_{\mp} ~=~\frac{[2\pm y]}{[1\pm y]^{2}}~\frac{\teff}{2}~,
\end{equation}
with $W_{-} \geq W_{+}$ (since $y =\tfrac{1}{2}t_{*}g_{c}(0) \in (-1,0)$ for all 
$\epsilon_{c}/t_{*} \in (1,\infty)$, eq.\ \ref{eq:127a}). For the case 
$\epsilon_{c}/t_{*} \gg 1$, where $y \rightarrow 0$ and $\lambda(\epsilon) \rightarrow \epsilon/\teff$ 
(eq.\ \ref{eq:141}), $W_{\mp}$ both approach $\teff$. The DoS $\tilde{\rho}(\epsilon)$ then asymptotically approaches
\begin{equation}
\label{eq:147}
\tilde{\rho}_{0}(\epsilon) ~=~\frac{2}{\pi\teff} \sqrt{1-(\epsilon/\teff)^{2}}~~~~~~:~ 
\frac{\epsilon_{c}}{t_{*}} \gg 1 
\end{equation}
-- a symmetric semicircular spectrum (\emph{cf} eq.\ \ref{eq:12}), but with a \emph{renormalised} band halfwidth, $\teff$, 
indicative of the purely NN Hubbard model arising in this limit.

\subsection{Qualitative consequences: $\epsilon_{c}/t_{*} \gg 1$}
\label{subsection:qualitativeobs}

Before pursuing the general mapping to a one-band long-ranged Hubbard model in the vicinity of a Mott transition, 
we consider some qualitative physical consequences of it for the regime $\epsilon_{c}/t_{*} \gg 1$ 
(sec.\ \ref{subsubsection:bigepsilonc}); as thus relevant close to the transition to a Mott-Hubbard  
insulator in the ZSA scheme~\cite{Zaanen-S-A1985} (sec.\ \ref{subsection: Phasediagtake2} and 
fig.\ \ref{fig:fig4}). Four points in particular should be noted:\\
\noindent {\bf{(i)\rm}}
Since $\gc$ is pure real, eqs.\ \ref{eq:128},\ref{eq:132} show that the $c$- and $f$-band spectra 
are asymptotically related by 
\begin{equation}
\label{eq:148}
D_{c}(\w) ~=~ \left(\frac{V}{\epsilon_{c}}\right)^{2}D_{f}(\w)~.
\end{equation}
Eq.\ \ref{eq:148} conforms as it must to the general result eq.\ \ref{eq:119} for the Fermi level  ($\w=0$) spectra 
at the Mott transition. But it holds generally for \emph{all} $\w \ll \epsilon_{c}-t_{*}$ 
(sec.\ \ref{subsubsection:bigepsilonc}); where the $\w$-dependences of $D_{c}(\w)$ and $D_{f}(\w)$ are thus the same,
with $D_{c}(\w)$ simply `cut down' from $D_{f}(\w)$ by the factor $(V/\epsilon_{c})^{2} \ll 1$.
This behaviour is indeed observed in NRG calculations; being apparent even e.g.\ for $\epsilon_{c}/t_{*} =3$, as shown
in fig.\ \ref{fig:fig1}.\\ 
\noindent {\bf{(ii)\rm}}
For $\epsilon_{c}/t_{*} \gg 1$, the entire free ($V=0$) $c$-band DoS -- $d_{0}^{c}(\w) =\rho_{0}(\w -\epsilon_{c})$,
centred on $\w =\epsilon_{c}$ -- lies far in excess of the Fermi level and as such is 
`empty'. This is naturally reflected in the full $c$-band spectrum $D_{c}(\w)$, where the great bulk of its 
spectral weight lies within $\sim \pm t_{*}$ of $\w =\epsilon_{c}$ (as seen e.g.\ in the inset to fig.\ \ref{fig:fig1}).
Eq.\ \ref{eq:148} nevertheless shows that, due to the local hybridization $V$ between the high-lying $c$-level and the 
$f$-level, the full $c$-band spectrum $D_{c}(\w)$  acquires a small spectral weight for energies 
$\w \ll \epsilon_{c}-t_{*}$ which encompass the Fermi level; thereby enabling a metallic state to persist in a regime where one might naively expect otherwise. \\
\noindent {\bf{(iii)\rm}}
As one anticipates near the transition to a Mott-Hubbard insulator,~\cite{Zaanen-S-A1985} the $f$- and $c$-bands in the general vicinity of the Fermi level are \emph{narrow}. The natural energy scale for these bands is not however the bare hopping $t_{*}$, but rather the effective hopping $\teff \sim t_{*}(V/\epsilon_{c})^{2} \ll t_{*}$.
This  too is evident in fig.\ \ref{fig:fig1}, for which $W_{\mathrm{eff}} = 2\teff \simeq 0.04 t_{*}$ (some two orders of magnitude less than the scale $W=2t_{*}$ associated with $d_{0}^{c}(\w)$).  Likewise, the low-energy `coherence' scale 
$\w_{\mathrm{L}}$ characteristic of the Fermi liquid Kondo resonance in the $f$- and $c$-electron spectra is not 
$\sim Zt_{*}$, but rather $\w_{\mathrm{L}}\sim Z \teff \ll Zt_{*}$  (providing a natural explanation for the marked difference observed between $Zt_{*}$ and the coherence scale in 
Ref.\ [\onlinecite{AmaricciPRB2012}]). \\
\noindent {\bf{(iv)\rm}}
Since eq.\ \ref{eq:148} holds in particular for all $\w$ around and below the Fermi level,  
the $c$- and $f$-level charges are  directly related by $n_{c}=(V/\epsilon_{c})^{2}n_{f}$. 
Hence, since $n_{t}=n_{c}+n_{f}=1$ at the Mott transition, 
\begin{equation}
\label{eq:149}
n_{f} ~=~ 1-\left(\frac{V}{\epsilon_{c}}\right)^{2}, ~~~~~~ n_{c} ~=~\left(\frac{V}{\epsilon_{c}}\right)^{2}
\end{equation} 
hold asymptotically for $\epsilon_{c}/t_{*} \gg 1$. This behaviour is confirmed by our NRG calculations. It also underscores the fact~\cite{SordiPRL2007,*SordiPRB2009,AmaricciPRB2012} that, for the transition to a Mott-Hubbard insulator 
($\Delta \gg U$ in the ZSA scheme~\cite{Zaanen-S-A1985}), the $f$-level charge itself is very close, but not exactly equal, to one.


\subsubsection{Kondo lattice model}
\label{subsubsection:KLM}

Following on from the above, we point out that while the physics captured here is inherent to the periodic Anderson model, it is quite different from that arising in a Kondo lattice model (KLM). In the latter case  $f$-level electrons are replaced by strictly localised spins, which are purely exchange-coupled to an otherwise free $c$-band, and hence generate scattering of 
$c$-band electrons. But if the free $c$-band is empty -- as arises e.g.\ for $\epsilon_{c} \gg t_{*}$ -- then there are no electrons to scatter, and the system is insulating. Indeed by this reasoning one expects the KLM to be insulating for any  
$\epsilon_{c}>t_{*}$; in marked contrast to the PAM (e.g.\ fig.\ \ref{fig:fig2}) where a metallic phase persists up to critical $\epsilon_{c}$'s far in excess of $t_{*}$. This reflects the fact that the procedure of 
secs.\ \ref{subsection:lowenergyHM},\ref{subsection:longrangeHM} generates an effective one-electron \emph{hopping}  
between $f$-level electrons in the PAM, via higher-order (and indeed fundamentally different) virtual processes to those 
arising in the conventional Schrieffer-Wolff transformation employed to map a PAM onto a KLM; and does so moreover without the constraint that $n_{f}$ is precisely unity (which itself would preclude a non-trivial Mott transition, for which
$n_{t}=n_{f}+n_{c} =1$ is required). Put more bluntly, the rich physics of the Mott transition in the PAM is simply absent in the KLM.


\subsection{Phase diagram}
\label{section:pdexplaining}

Given the mapping to an effective one-band Hubbard model in the vicinity of a Mott transition, we 
revisit the phase diagram of the PAM in the light of it. As in previous sections we naturally focus on the 
Mott transition for $n_{t}=1$ (requiring as such $\epsilon_{c}/t_{*} >1$ and $\eta_{f} \geq 0$).


\subsubsection{Phase boundary: scaling collapse}
\label{subsubsection:pbcollapse}

The effective one-band Hubbard model onto which the PAM maps in the vicinity of the Mott transition, is of course determined by \emph{its} three intrinsic parameters: (a) $U/\teff$; (b) $y = \tfrac{1}{2}t_{*}g_{c}(0)$ (eq.\ \ref{eq:127a}),
which controls the ratio of the $n^{\mathrm{th}}$ NN  to the NN hopping amplitudes $\teff^{(n)}/\teff$ (eq.\ \ref{eq:144});
and (c)  the asymmetry 
\begin{equation}
\label{eq:150}
\neff ~:=~ 1~+~\frac{2\epsilon_{0}}{U},
\end{equation}
given by (see  eq.\ \ref{eq:145b}) 
\begin{equation}
\label{eq:151}
\neff ~=~ \eta_{f}^{\pd}~+~\frac{\teff}{U y}
\end{equation}
with $\eta_{f}$ the $f$-level asymmetry (eq.\ \ref{eq:13}).

\begin{figure}
\includegraphics{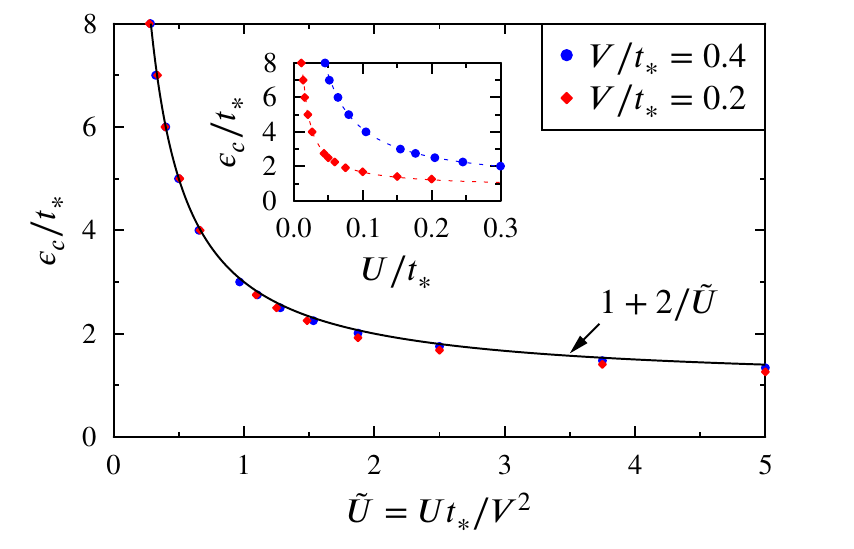}
\caption{\label{fig:fig5} 
NRG results for the PAM Mott transition, for fixed $\eta_{f}=0$,
shown for two different hybridization strengths $V/t_{*} = 0.4$ (circles) and $0.2$ (diamonds). Inset
shows the critical $\epsilon_{c}/t_{*}$ \emph{vs} $U/t_{*}$, while the main figure shows $\epsilon_{c}/t_{*}$
\emph{vs} $\tilde{U}=Ut_{*}/V^{2}$. When scaled in this way the phase boundary is 
seen to be independent of the hybridization (sec.\ \ref{subsubsection:pbcollapse}) to within numerical accuracy.  
The figure also shows comparison to eq.\ \ref{eq:154} (solid line); which is asymptotically exact for 
$\epsilon_{c}/t_{*} \gg 1$, but captures the phase boundary very well over essentially the full $\epsilon_{c}/t_{*}$ range.
}
\end{figure}

Notice then that for given $\eta_{f}$, the asymmetry $\neff$ is entirely determined by $U/\teff$ and $y$; whence the effective Hubbard model depends solely on  $U/\teff$ and $y$ ($\equiv y(\epsilon_{c}/t_{*})$). 
This is one parameter fewer than for the PAM itself, which for given $\eta_{f}$ is specified by $U/t_{*}$, 
$\epsilon_{c}/t_{*}$ and $V/t_{*}$; and this parametric redundancy clearly implies a scaling collapse of the phase boundary for the Mott transition. The origin of this is clear from eq.\ \ref{eq:139}: $V$ enters the effective Hubbard model parameters solely through $\teff \propto V^{2}/t_{*}$.
Hence the critical $U/\teff$ as a function of $\epsilon_{c}/t_{*}$ (which determines $y$) must be \emph{independent} 
of the hybridization; or equivalently $\epsilon_{c}/t_{*}$ \emph{vs} $Ut_{*}/V^{2}$ must be $V$-independent.
NRG results for the M/MI transition in the PAM are shown in fig.\ \ref{fig:fig5} for two different values of $V/t_{*}$,
with the inset showing the phase boundaries as $\epsilon_{c}/t_{*}$ \emph{vs} $U/t_{*}$. 
The resultant scaling collapse is seen in the main figure, showing the same data \emph{vs} $\tilde{U}=Ut_{*}/V^{2}$
(and we attribute the small deviation from perfect collapse to numerical inaccuracies in our NRG calculations).


\subsubsection{Mott transition phase boundaries}
\label{subsubsection:pbsimple}

To gain further insight we consider now a physically intuitive, if in general approximate, estimate for the
Mott transition boundaries, by simple consideration of the spectrum $D(\w)$ of the underlying effective one-band Hubbard model.
The already apparent richness of the phase behaviour will be seen to reflect the fact that the effective model has 
long-ranged (rather than solely NN) hoppings, as embodied in the non-trvial $y \equiv y(\epsilon_{c}/t_{*})$-dependence of the problem.

 $D(\w)$ in the MI phase consists of lower and upper Hubbard bands (HBs) centred on $\w =\epsilon_{0}$ and 
$\w =\epsilon_{0}+U$ respectively (with site-energy $\epsilon_{0}$  given by eq.\ \ref{eq:145}). Model the HBs simply as 
non-interacting bands centred on $\w =\epsilon_{0}$, $\epsilon_{0}+U$. The upper edge of the lower Hubbard band (LHB) then occurs at $\w =\epsilon_{0}+W_{+}$, and the lower edge of the upper Hubbard band (UHB) at $\w =\epsilon_{0}+U-W_{-}$; with 
$W_{\mp}$ given by eq.\ \ref{eq:146}. Now adopt the familiar simple estimate that the system is a MI provided the Fermi level $E_{F} \equiv 0$ lies above the upper edge of the LHB and below the lower edge of the UHB; i.e.\ provided
$\epsilon_{0} +W_{+} \leq 0 \leq \epsilon_{0}+U-W_{-}$. Recalling that $\epsilon_{f} = -\tfrac{1}{2}U(1-\eta_{f})$, and
using eq.\ \ref{eq:139a} for $\teff$, this condition is readily shown to reduce to
\begin{equation}
\label{eq:152}
\tilde{U}(\eta_{f}-1) - \frac{2}{(\frac{\epsilon_{c}}{t_{*}}+1)} ~\leq ~ 0 ~\leq ~\tilde{U}(\eta_{f}+1) - 
\frac{2}{(\frac{\epsilon_{c}}{t_{*}}-1)}
\end{equation}
where
\begin{equation}
\label{eq:153}
\tilde{U}~=~ \frac{Ut_{*}}{V^{2}}
\end{equation}
is thus defined (such that the entire $V$-dependence of the phase boundary is encoded on $\tilde{U}$, as 
established on general grounds in sec.\ \ref{subsubsection:pbcollapse} above). 

The right inequality in eq.\ \ref{eq:152} is the condition for the lower edge of the UHB to lie above or precisely at
the Fermi level. The latter corresponds to the $n_{t} = 1+$ (`electron-doped') Mott transition, the critical
$\epsilon_{c}$ for which (denoted by $\epsilon_{c_{>}}$) is thus given by
\begin{equation}
\label{eq:154}
\frac{\epsilon_{c_{>}}}{t_{*}}~=~1~+~ \frac{2}{(\eta_{f}+1)~\tilde{U}}~~~~~~:~n_{t}= 1+
\end{equation}
(holding for the full range $\eta_{f} \geq 0$).
The left inequality in eq.\ \ref{eq:152} is by contrast the condition for the upper edge of the LHB to lie below or 
at the Fermi level, the latter corresponding to an $n_{t} = 1-$ (`hole-doped') Mott transition. But for 
$\eta_{f} \leq 1$, the left hand side of this inequality is necessarily negative for all $U\geq 0$. An $n_{t} = 1-$ 
Mott transition cannot therefore occur for $\eta_{f} <1$, as argued generally in sec.\ \ref{subsection: Phasediagtake2} and confirmed by NRG (fig.\ \ref{fig:fig3}).

For $\eta_{f} >1$, both $n_{t} = 1-$ and $n_{t} = 1+$ transitions can then occur. The critical 
$\epsilon_{c} \equiv \epsilon_{c_{>}}$ for $n_{t} = 1+$  is again given by eq.\ \ref{eq:154}, while 
$\epsilon_{c} \equiv \epsilon_{c_{<}}$ for the $n_{t} = 1-$ Mott transition follows from the left side equality in 
eq.\ \ref{eq:152} as 
\begin{equation}
\label{eq:155}
\frac{\epsilon_{c_{<}}}{t_{*}}~=~-1~+~ \frac{2}{(\eta_{f}-1)~\tilde{U}}~~~~~~:~n_{t}= 1-~.
\end{equation}
The Mott insulating region in the $(\tilde{U}, \epsilon_{c}/t_{*})$-plane is thus bounded by 
the lines $\epsilon_{c_{>}}$ and $\epsilon_{c_{<}}$ $(\geq \epsilon_{c_{>}})$; which, 
since eq.\ \ref{eq:152} must be satisfied, requires 
$\epsilon_{c} \geq \epsilon_{c_{*}}$ and $\tilde{U} \leq \tilde{U}_{*}$, given (on equating 
$\epsilon_{c_{>}}=\epsilon_{c_{<}}$, eqs.\ \ref{eq:154},\ref{eq:155})
by
\begin{equation}
\label{eq:156}
\frac{\epsilon_{c_{*}}}{t_{*}} ~=~ \eta_{f}, ~~~~
\tilde{U}_{*}~=~\frac{2}{(\eta_{f}^{2}-1)}.
\end{equation}

The asymmetry of the effective Hubbard model follows from eqs.\ \ref{eq:151},\ref{eq:139a},\ref{eq:153} as
$\neff = \eta_{f} -4|y|/\tilde{U}$; whence the asymmetries for the two Mott transition  boundaries (denoted by
$\eta_{\mathrm{eff}_{\gtrless}}$) are given using eqs.\ \ref{eq:154},\ref{eq:155} by
\begin{equation}
\label{eq:157}
\eta_{\mathrm{eff}_{\gtrless}} ~=~ \eta_{f} ~-~ 2(\eta_{f}\pm1)(x\mp 1) \left[x -\sqrt{x^{2}-1}\right]
\end{equation}
where $x =\epsilon_{c}/t_{*}$. For $\epsilon_{c}/t_{*}\gg 1$ in particular, this yields
\begin{equation}
\label{eq:158}
\eta_{\mathrm{eff}_{\gtrless}} ~\overset{\epsilon_{c}/t_{*}\gg 1}{\sim} ~\mp 1  
\end{equation}
independently of $\eta_{f}$. The $\neff$'s for the $n_{t} = \pm 1$ Mott transitions are in otherwords the maximum 
that are possible~\cite{DELMRG2016} for a purely NN Hubbard model (which arises
asymptotically for $\epsilon_{c}/t_{*} \gg 1$, sec.\ \ref{subsubsection:bigepsilonc});
and in which regime the metallic phase Kondo resonance in $D(\w)$ indeed sits at the edges of the 
gap~\cite{DELMRG2016} directly adjacent to the Hubbard bands. For this reason, we expect the above 
estimates for the Mott transition phase boundaries to be asymptotically exact for $\epsilon_{c}/t_{*} \gg 1$.
For $x =\epsilon_{c}/t_{*} \rightarrow 1+$ by contrast (relevant only to $\eta_{\mathrm{eff}_{>}}$ as explained above),
eq.\ \ref{eq:157} gives $\eta_{\mathrm{eff}_{>}}\rightarrow \eta_{f}$, and the asymmetries of the effective Hubbard model 
and the PAM itself then coincide. 

\begin{figure}
\includegraphics{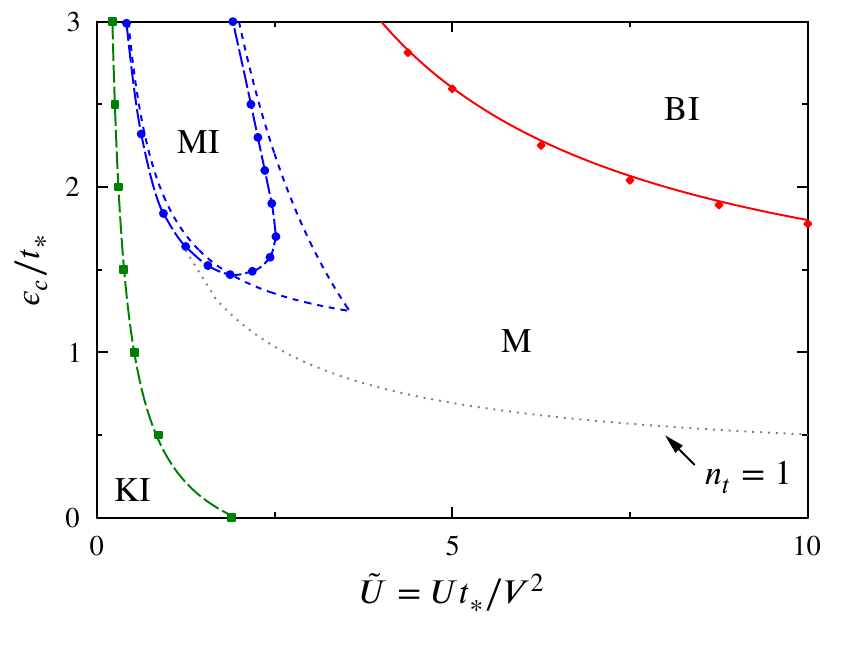}
\caption{\label{fig:fig6} 
NRG results for PAM phase boundaries in the ($\tilde{U}, \epsilon_{c}/t_{*}$)-plane 
($\tilde{U} =Ut_{*}/V^{2}$), for fixed $\eta_{f}=1.25$ (with long-dash line as guide to eye). Phases shown are metal, Mott insulator (MI, $n_{t}=1$), band insulator (BI, $n_{t}=0$), Kondo insulator (KI, $n_{t}=2$). Both $n_{t} =1+$ (electron-doped) and $n_{t} =1-$ (hole-doped) Mott transitions occur for $\eta_{f} >1$, as discussed in text 
(and the dotted line simply shows the location of $n_{t}=1$ in the M phase).
The estimates eqs.\ \ref{eq:154},\ref{eq:155} for the two Mott boundaries are shown (short-dash lines), and become asymptotically exact for $\epsilon_{c}/t_{*} \gg 1$. The exact result eq.\ \ref{eq:161} for the metal/BI phase boundary is also indicated (solid line); NRG points are barely distinguishable from it.
}
\end{figure}

 The simple predictions above are compared to NRG calculations in figs.\ \ref{fig:fig5},\ref{fig:fig6}.
Eq.\ \ref{eq:154} is compared in fig.\ \ref{fig:fig5} to the NRG-determined phase boundary for $\eta_{f} =0$.
While we expect it to be asymptotically exact for $\epsilon_{c_{>}}/t_{*} \gg 1$, it 
clearly describes very well the phase boundary over essentially the full range; reflecting no doubt the
fact that it also captures correctly the  $\tilde{U} \rightarrow \infty$ asymptote
$\epsilon_{c_{>}}=t_{*}$ (sec.\ \ref{subsection: Phasediag}), and that the metallic Kondo resonance in practice 
lies close to the lower edge of the UHB even for fairly small $\epsilon_{c}/t_{*}$, see  e.g.\ fig.\ \ref{fig:fig1}.

Fig.\ \ref{fig:fig6} shows NRG results for $\eta_{f} = 1.25$, and contains the full range of phases -- metal, Mott insulator, band insulator, Kondo insulator. The simple estimates eqs.\ \ref{eq:154},\ref{eq:155} for the two Mott insulator boundaries are indicated (short-dashed lines). While not wholly quantitative, particularly in the vicinity 
of  ($\tilde{U}_{*},\epsilon_{c*}$) ($\simeq (3.56,1.25)$ here), they nevertheless capture
the behaviour well and, as anticipated, appear to become asymptotically exact for $\epsilon_{c}/t_{*} \gg 1$.\\

Finally, we return to a point noted in sec.\ \ref{subsection: Phasediag}, and evident e.g.\ in 
figs.\ \ref{fig:fig2},\ref{fig:fig3}: for $\epsilon_{c}/t_{*} \gg 1$, the critical $U/t_{*}$(s) for the Mott transition asymptotically vanish, and appear as such to be in `weak coupling'.
From eqs.\ \ref{eq:154},\ref{eq:155} for example, the critical $U/t_{*}$ for the two Mott transitions (call them
$U_{\gtrless}/t_{*}$) follow as
\begin{equation}
\label{eq:159}
\frac{U_{\gtrless}}{t_{*}} ~=~ \left(\frac{V}{t_{*}}\right)^{2} 
\frac{2}{(\eta_{f}\pm 1)} ~\frac{1}{[\frac{\epsilon_{c}}{t_{*}} \mp 1]}~~~~~~:~n_{t}=1\pm
\end{equation}
and for  $\epsilon_{c}/t_{*} \gg 1$ indeed vanish asymptotically. While correct, as pointed out in 
sec.\ \ref{subsection:qualitativeobs} the relevant scale with which to compare the interaction $U$ 
is the \emph{effective} hopping $\teff$, which likewise vanishes asymptotically for  
$\epsilon_{c}/t_{*} \gg 1$ (eq.\ \ref{eq:134}). Expressed in these terms, the asymptotic behaviour is clearly
\begin{equation}
\label{eq:160}
\frac{U_{\gtrless}}{\teff} ~=~\frac{2}{(\eta_{f}\pm 1)} ~\frac{\epsilon_{c}}{t_{*}}~~~~~~~~:~\frac{\epsilon_{c}}{t_{*}}\gg 1
\end{equation}
and thus \emph{grows} with increasing $\epsilon_{c}/t_{*}$, indicative of the strongly correlated nature of the Mott transition.


\subsubsection{Metal/band insulator}
\label{subsubsection:BIboundary}

The phase boundary between the metal and the band insulator can also be obtained simply and exactly. Recall that as a result of Luttinger's theorem the total charge $n_{t}$ is given by eq.\ \ref{eq:117}, with 
$\epsilon_{f}^{*} =\epsilon_{f}+\Sigma^{R}_{f}(\w =0)$ the renormalised level.
The boundary to the $n_{t}=0$ band insulator obviously  requires $\epsilon_{f}^{*} >0$, 
and  corresponds to $-\epsilon_{c}+V^{2}/\epsilon_{f}^{*} = -t_{*}$  (i.e.\ to the lower band edge of $\rho_{0}(\epsilon)$
in eq.\ \ref{eq:117}).  But for $n_{t}=0$ there are no electrons in the system and the problem is thus effectively
 non-interacting. Hence $\epsilon_{f}^{*} \equiv \epsilon_{f}$, 
so the boundary to the band insulator is $\epsilon_{c}/t_{*} = 1+V^{2}/(t_{*}\epsilon_{f})$. 
But $\epsilon_{f} = \tfrac{1}{2}U(\eta_{f}-1)$ (and the requirement $\epsilon_{f}^{*} =\epsilon_{f}>0$ is equivalently just that $\eta_{f} >1$ is required for the metal to $n_{t}=0$ band insulator to arise).
Noting eq.\ \ref{eq:153} defining $\tilde{U}$, the phase boundary for the metal to band insulator is thus
\begin{equation}
\label{eq:161}
\frac{\epsilon_{c}}{t_{*}} ~=~ 1~+~ \frac{2}{(\eta_{f}-1)~\tilde{U}} ~~~~~~:~n_{t}=0~.
\end{equation}
This is also compared to NRG results for $\eta_{f}=1.25$ in fig.\ \ref{fig:fig6}, which  are in essence indistinguishable from it.


\subsection{Single-particle dynamics}
\label{section:spdynamics}
 
We touch now on single-particle dynamics in the metal close to the Mott transition, to illustrate 
typical behaviour associated with the two classes of insulators in the ZSA~\cite{Zaanen-S-A1985} scheme 
(sec.\ \ref{subsection: Phasediagtake2}) -- Mott-Hubbard and charge-transfer insulators.
We do so with reference to the phase diagram fig.\ \ref{fig:fig4} (for $\eta_{f}=0$), where point A is representative of  behaviour close to a MH insulator, and point B to a CT insulator.

\begin{figure}
\includegraphics{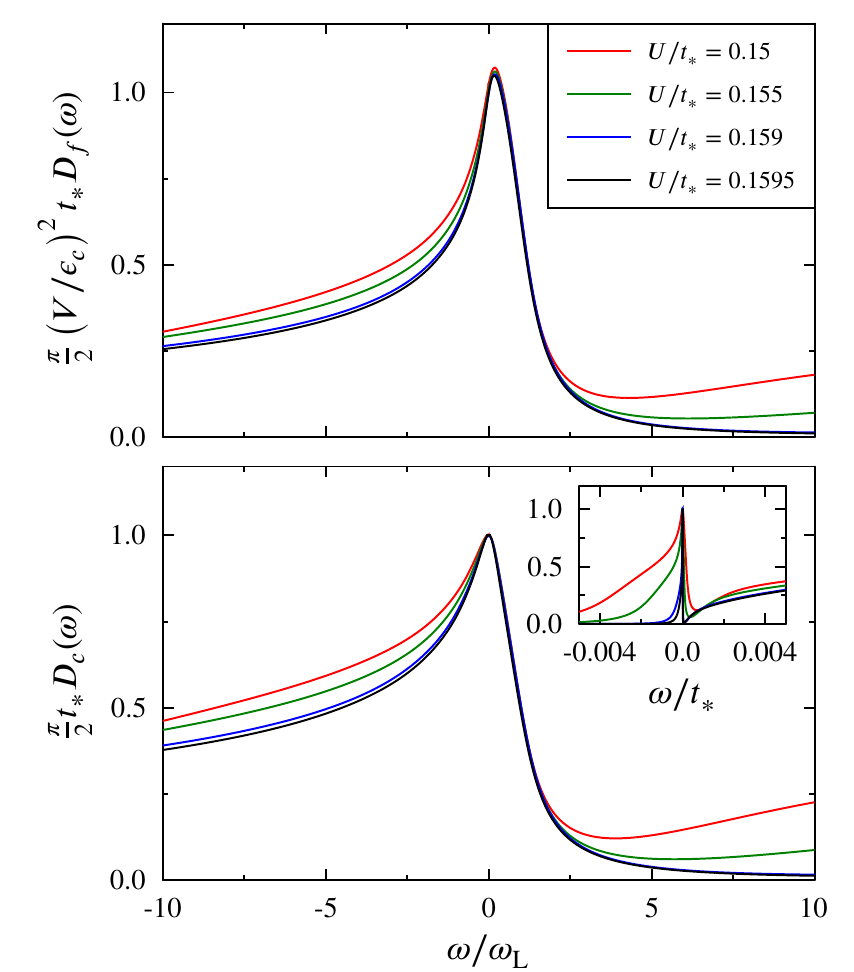}
\caption{\label{fig:fig7} 
Scaling of dynamics in the metal close to the transition to a Mott-Hubbard insulator 
(see point A in fig.\ \ref{fig:fig4} and fig.\ \ref{fig:fig1}). Shown for the 
same parameters as fig.\ \ref{fig:fig1}, with $U/t_{*}$'s as indicated.
As the transition is approached and the low-energy scale $\w_{_{\mathrm{L}}}$ vanishes, the spectra 
exhibit scaling collapse as a function of $\w/\w_{_{\mathrm{L}}}$.~\cite{omegaLdef}
\emph{Upper panel}: $f$-band spectrum $\tfrac{\pi}{2}t_{*}(V/\epsilon_{c})^{2}D_{f}(\w)$ \emph{vs} 
$\w/\w_{_{\mathrm{L}}}$.
\emph{Lower panel}: $c$-band spectrum $\tfrac{\pi}{2}t_{*}D_{c}(\w)$ \emph{vs} $\w/\w_{_{\mathrm{L}}}$.
Inset: $c$-spectrum 
\emph{vs} $\w/t_{*}$.
}
\end{figure}

 Single-particle dynamics in the vicinity of point A have been shown in fig.\ \ref{fig:fig1}, and their qualitative characteristics discussed in sec.\ \ref{subsection:qualitativeobs} (and sec.\ \ref{subsection:FLphases}).
The great bulk of the $c$-band spectrum lies far above the Fermi level (fig.\ \ref{fig:fig1} inset),
and is effectively a passive spectator to the Mott transition. The `action' in $D_{c}(\w)$ around the Fermi level is 
instead generated by hybridization $V$ to the high-lying $c$-level, leading to characteristically narrow Hubbard bands with widths controlled by the effective hopping $\teff \ll t_{*}$. The low-energy Kondo resonance -- occurring in both $D_{c}(\w)$ and $D_{f}(\w)$ -- lies at the lower edge of the upper Hubbard band (fig.\ \ref{fig:fig1}), narrows progressively as the transition is approached and its Kondo scale $\w_{\mathrm{L}}$ vanishes; and collapses on the spot as the transition is crossed, to yield the MH insulator with a finite spectral gap of order $\Delta_{g} \simeq U -2\teff$. 
Since $\w_{\mathrm{L}}$ vanishes continuously on approaching the transition from the metal, the spectra exhibit scaling as a function of $\w/\w_{\mathrm{L}}$; as shown explicitly in fig.\ \ref{fig:fig7} for both the $c$- and 
$f$-spectra,~\cite{omegaLdef}
which are seen to have rather asymmetric scaling resonances.

\begin{figure}
\includegraphics{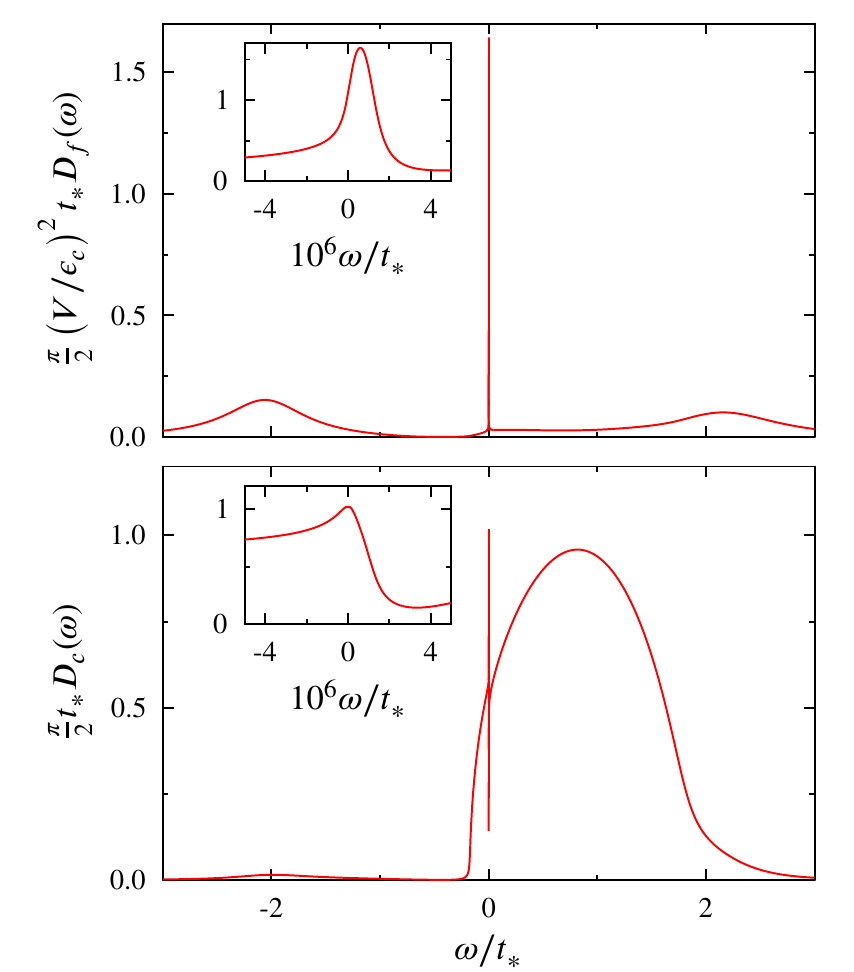}
\caption{\label{fig:fig8} 
Typical single-particle dynamics 
close to the transition to a charge-transfer insulator (point B in fig.\ \ref{fig:fig4}: $\epsilon_{c}/t_{*} =0.85$,  
$U/t_{*}=4$, with $\epsilon_{f}/t_{*} =-2$ since $\eta_{f}=0$). 
\emph{Upper panel}: $f$-spectrum shown as $\tfrac{\pi}{2}t_{*}(V/\epsilon_{c})^{2}D_{f}(\w)$ \emph{vs} $\w/t_{*}$. 
\emph{Lower panel}: $c$-band spectrum $\tfrac{\pi}{2}t_{*}D_{c}(\w)$ \emph{vs} $\w/t_{*}$. 
\emph{Insets}: spectra on low-energy scales; since ${\w}_{_{\mathrm{L}}} \simeq 10^{-6}t_{*}$ here, these amount in practice to the scaling forms of the spectra. See text for discussion.
}
\end{figure}

 Typical dynamics close to the Mott transition to a CT insulator are illustrated in fig.\ \ref{fig:fig8} (for point B of
fig.\ \ref{fig:fig4}, $\epsilon_{c}/t_{*} =0.85$ and $U/t_{*}=4$); again showing $c$- and $f$-level spectra as
$\tfrac{\pi}{2}t_{*}D_{c}(\w)$ and $\tfrac{\pi}{2}t_{*}(V/\epsilon_{c})^{2}D_{f}(\w)$ \emph{vs} $\w/t_{*}$. They have 
clear differences from, but important similarities to, those close to a MH insulator (fig.\ \ref{fig:fig1}).
The relevant bands are now by contrast broad, with their widths dictated by the bare hopping $t_{*}$; and -- in marked contrast to eq.\ \ref{eq:148} close to a MH insulator -- the $\w$-dependences of the $c$- and $f$-spectra are quite different.
Hubbard bands are for example evident in the $f$-spectrum (centred around $\w \simeq \pm \tfrac{1}{2}U$ $=2t_{*}$), but are barely apparent in the $c$-spectrum, which is instead dominated by the nearly-free semicircular spectrum centred on 
$\w \simeq \epsilon_{c}$ with halfwidth $\sim t_{*}$.

The physics at low-energies around the Fermi level remains however characterised by the 
Kondo-like resonances evident in both spectra. These again lie close to the lower edge of the upper Hubbard band 
in $D_{f}(\w)$, narrow progressively as the low-energy scale $\w_{_{\mathrm{L}}}$ vanishes  on approaching the transition 
(e.g.\ by increasing $\epsilon_{c}$ towards its critical value $\sim t_{*}$); and again collapse on the spot as the transition is crossed, to yield the CT insulator whose finite spectral gap $\Delta_{g}$ is now controlled by the charge transfer energy $\Delta =\epsilon_{c}-\epsilon_{f}$ (rather than by $U$), and is of order 
$\Delta_{g} \simeq  \Delta -2t_{*}$ ($\simeq t_{*}$ for the case considered).

The insets to fig.\ \ref{fig:fig8} show the low-energy Kondo resonances of both $f$- and $c$-level spectra, for 
$\epsilon_{c}/t_{*} =0.85$ in the metal close to the transition, where $n_{t}$ is very close to one and the low-energy scale 
$\w_{_{\mathrm{L}}} \simeq 10^{-6}t_{*}$ (sufficiently small that the resonances shown amount in practice to their scaling forms).  The Fermi level spectra indeed accurately satisfy eq.\ \ref{eq:119}, but the resonances in the $f$- and $c$-spectra are clearly different; that for the $f$-level being relatively weakly asymmetric (with its maximum displaced slightly above the Fermi level to $\w \simeq \w_{_{\mathrm{L}}}$), while that for the $c$-level 
is markedly asymmetric.

Finally, we add that single-particle spectra for point C of fig.\ \ref{fig:fig4} -- corresponding to the crossover region where the charge transfer energy $\Delta \simeq U$ -- are qualitatively similar to those shown in fig.\ \ref{fig:fig8} close to a CT insulator; and for that reason are omitted here.


\section{Mott insulator}
\label{section:MottPhase}
Thus far we have said  
little about the Mott insulator phase itself.
We turn to it now.

In the Fermi liquid phases the electron spin degrees of freedom are completely quenched, reflected in a non-degenerate ground state with e.g.\ a vanishing $T=0$ entropy. By contrast, the MI phase within DMFT --  be it for the Hubbard model or the PAM -- is characterised by a residual entropy of $k_{B}\ln2$ per site;~\cite{dmftkotliar} reflecting incomplete Kondo-like quenching of electron spins, and hence an unquenched local moment per site (denoted by $\tilde{\mu}$).

To handle the locally doubly-degenerate MI phase requires a two-self-energy (TSE) description, 
recently considered in the context of both a range of quantum impurity models~\cite{LTG2014} and the one-band Hubbard model 
within DMFT.~\cite{DELMRG2016} We draw on this work in the following, and confine ourselves below to a brief summary of key 
features; emphasising that the TSE description used here is exact 
(it also underlies the local moment approach~\cite{LET,*LMALETHubbard1996,*MTGLMA_asym,*nigelscalspec}$^{,}$\cite{RajaEPJB,vickiSPAMEPJB2003,KIPAMJPCM2003,rajapamtheory,*rajapamexp}
but its use there, while providing a rather successful description of e.g.\ the 
PAM,~\cite{RajaEPJB,vickiSPAMEPJB2003,KIPAMJPCM2003,rajapamtheory,*rajapamexp,Parihari+NSV2008,NSV+KumarJPCM2011}  
is in general approximate).

Within a TSE description,\cite{LTG2014,DELMRG2016} the local propagators $G_{\nu}(\w)$ 
with $\nu =c,f$, are expressed as $G_{\nu}(\w)=\tfrac{1}{2}[G_{\nu_{A\sigma}}(\w)+G_{\nu_{B\sigma}}(\w)]$. 
$G_{\nu_{A\sigma}}(\w)$ refers to the propagator for local moment $\tilde{\mu}=+\mutil$,
while $G_{\nu_{B\sigma}}(\w)$ refers to that for $\tilde{\mu}=-\mutil$.~\cite{momentsign}
From the invariance of $H$ under spin exchange it follows that $G_{\nu_{A\sigma}}(\w)=G_{\nu_{B-\sigma}}(\w)$, such
that $G_{\nu}(\w)$ is correctly rotationally invariant (independent of $\sigma$). Hence
\begin{equation}
\label{eq:162}
G_{\nu}(\w) ~=~ \tfrac{1}{2}\left[ G_{\nu _{A\uparrow}}(\w)~+~G_{\nu _{A\downarrow}}(\w)\right]
~~~~:~\nu = c,f
\end{equation}
enabling one to consider solely the $A$-type propagators, as employed in the following. The  
$G_{\nu_{A\sigma}}(\w)$ are given in terms of the corresponding two-self-energies 
$\Sigma_{f_{A\sigma}}(\w)$ ($=\Sigma_{f_{A\sigma}}^{R}(\w)-i\Sigma_{f_{A\sigma}}^{I}(\w)$) by
(\emph{cf} eqs.\ \ref{eq:18})
\begin{subequations}
\label{eq:163}
\begin{align}
G_{c_{A\sigma}}(\w)~=&~ \Big[
\w^{+} -\epsilon_{c}^{\pd}- \frac{V^{2}}{\w^{+}-\epsilon_{f}^{\pd}  -\Sigma_{f_{A\sigma}}^{\pd}(\w)}-
\tfrac{1}{4}t_{*}^{2}G_{c}(\w)
\Big]^{-1}
\label{eq:163a}
\\
G_{f_{A\sigma}}(\w)=
&~\left[
\w^{+} -\epsilon_{f}^{\pd}- \Sigma_{f_{A\sigma}}^{\pd}(\w)-S_{f}^{\pd}(\w)
\right]^{-1},
\label{eq:163b}
\end{align}
\end{subequations}
with the local $c$- and $f$-level Feenberg self-energies again given precisely as in eq.\ \ref{eq:18} by 
$S_{c}(\w) =\tfrac{1}{4}t_{*}^{2}G_{c}(\w)$  and $S_{f}(\w)=V^{2}[\w^{+}-\epsilon_{c} -\tfrac{1}{4}t_{*}^{2}G_{c}(\w)]^{-1}$ (reflecting the fact that the nearest neighbours to any site are equally probably $A$-type ($\tilde{\mu}=+\mutil$) or 
$B$-type ($\tilde{\mu}=-\mutil$)).

As detailed in Ref.\ \onlinecite{LTG2014} it is the two-self-energies 
$\Sigma_{f_{A\sigma}}(\w)$ (and \emph{not} the conventional single self-energy $\Sigma_{f}(\w)$) that are directly calculable from many-body perturbation theory in the degenerate MI, as functional derivatives of a Luttinger-Ward functional. In consequence, a Luttinger theorem holds for the two-self-energies and associated propagators, namely
\begin{equation}
\label{eq:164}
I_{L_{A\sigma}}~=~\mathrm{Im}\int_{-\infty}^{0}d\w~G_{f_{A\sigma}}(\w)\frac{\partial\Sigma_{f_{A\sigma}}(\w)}{\partial \w}
~=0~.
\end{equation}
The standard Luttinger theorem applicable to the Fermi liquid phase, $I_{L}=0$ (with $I_{L}$ given by eq.\ \ref{eq:118} in terms of the single self-energy $\Sigma_{f}(\w)$), does not by contrast hold in the MI phase; we consider and determine it explicitly in sec.\ \ref{section:Luttint}. $\Sigma_{f}(\w)$ nonetheless remains defined in the MI just as in 
eq.\ \ref{eq:18}; comparison of which with eqs.\ \ref{eq:162},\ref{eq:163}  gives an exact relation between $\Sigma_{f}$ and the two-self-energies $\{\Sigma_{f_{A\sigma}} \}$
(which we omit here because no essential use is made of it in the following). Finally, as discussed in 
Refs.\ \onlinecite{LTG2014},\onlinecite{DELMRG2016}, we add that both the
$\Sigma_{f_{A\sigma}}(\w)$ and $\Sigma_{f}(\w)$ may be calculated using NRG (as will be employed below).


\subsection{Total site charge and moment}
\label{subsection:C&MMI}

The charges $n_{\nu}$ are as ever given from eq.\ \ref{eq:14} (with $G_{\nu}(\w)$ in the MI from
eq.\ \ref{eq:162}); with the total local charge $n_{t}=n_{c}+n_{f}$, and local moment $\mutil$, 
given by
\begin{subequations}
\label{eq:165}
\begin{align}
n_{t}=& \sum_{\nu =c,f} \tfrac{(-1)}{\pi}\mathrm{Im}
\int_{-\infty}^{0}d\w \left[ G_{\nu _{A\uparrow}}(\w)+G_{\nu _{A\downarrow}}(\w)\right]
\label{eq:165a}
\\
\mutil =&\sum_{\nu =c,f} \tfrac{(-1)}{\pi}\mathrm{Im}
\int_{-\infty}^{0}d\w \left[ G_{\nu _{A\uparrow}}(\w)-G_{\nu _{A\downarrow}}(\w)\right].
\label{eq:165b}
\end{align}
\end{subequations}
In parallel to $n_{t}$, the moment $\mutil$ is a `total' site local moment, in general receiving contributions from both 
$f$- and $c$-levels (naturally so, given the local hybridization $V$ coupling the levels). We focus on  the total site 
charge/moment -- rather than the separate contributions from $f$- and $c$-levels -- because it is these 
about which exact statements can be made, as below; this is directly analogous to the fact that 
eq.\ \ref{eq:117} for Fermi liquid phases, which has played an important role in our analysis of the metal,
refers likewise to the total local charge $n_{t}$.

\begin{figure}
\includegraphics{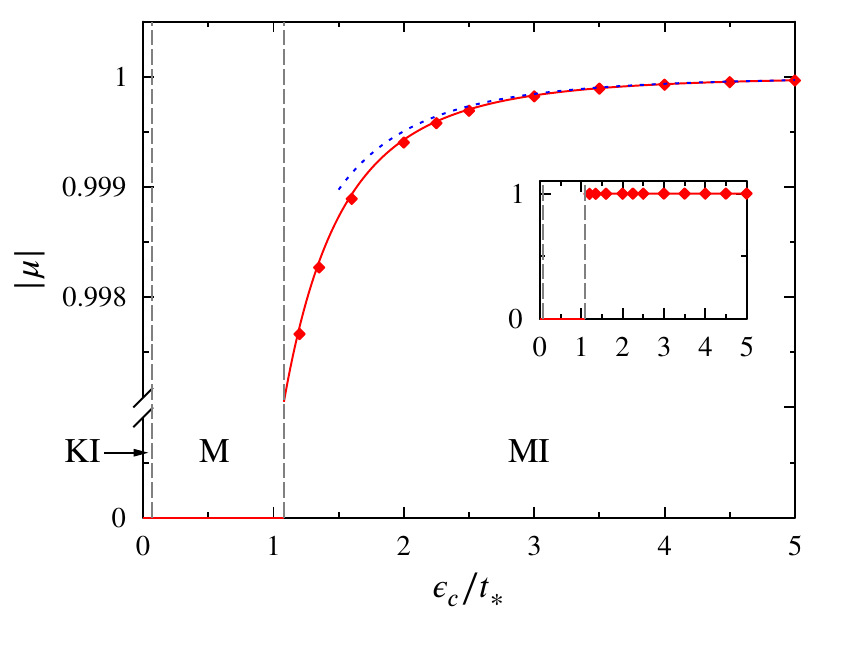}
\caption{\label{fig:fig9} 
NRG results (diamonds) for the local moment $\mutil$ \emph{vs} $\epsilon_{c}/t_{*}$, shown for fixed interaction 
$U/t_{*} = 2$ with 
$\eta_{f}=0$ and $V/t_{*} =0.4$. 
The moment is non-zero throughout the MI phase (arising for $\epsilon_{c}/t_{*} \gtrsim 1.1$, fig.\ \ref{fig:fig4}), 
vanishes discontinuously at the Mott transition (inset), and is zero in both the (Fermi liquid) M and KI phases.
Throughout the MI $\mutil$ deviates from, but remains remarkably close to, its saturation value of $1$ (main figure) for reasons explained in text. The leading asymptotic behaviour of 
$\mutil$ given by eq.\ \ref{eq:173} is also shown (solid line), and is seen to capture the NRG results; 
its ultimate asymptotic form, eq.\ \ref{eq:176}, is also shown for comparison (dotted line).
}
\end{figure}

While the charge is of course fixed at $n_{t}=1$ or $3$ throughout the MI phase, the local moment $\mutil$ is not. 
By way of example, fig.\ \ref{fig:fig9} shows typical NRG results for $\mutil$ as a function of $\epsilon_{c}/t_{*}$, for  a fixed interaction $U/t_{*}=2$ (with $\eta_{f} =0$, $V/t_{*} =0.4$); taking as such a 
cut through the phase diagram of fig.\ \ref{fig:fig4}. As occurs also for the usual one-band Hubbard 
model,~\cite{DELMRG2016} the moment is non-zero throughout the MI, and vanishes discontinuously at the Mott transition
(reflecting the fact that the critical line $U \equiv U_{c}(\epsilon_{c})$ for the transition is a line of
``$U_{c2}$'s" in standard terminology~\cite{dmftkotliar}). Local moments are in fact 
extremely well-developed in the insulator, for although $\mutil$ is seen to deviate steadily from its saturation value 
of $1$, it  remains strikingly close to it throughout the MI phase -- much more so than in the one-band Hubbard model
(e.g.\ fig.\ 2 of Ref.\ \onlinecite{DELMRG2016}).  An explanation for this behaviour is given below, see eqs.\ \ref{eq:173},\ref{eq:176}.

The essential results for $n_{t}$ and $\mutil$ are expressed in terms of interaction-renormalised levels associated with the 
two-self-energies, defined by (\emph{cf} eq.\ \ref{eq:113} for the Fermi liquid phases)
\begin{equation}
\label{eq:166}
\epsilon_{f_{A\sigma}}^{*}~=~\epsilon_{f}~+~\Sigma_{f_{A\sigma}}^{R}(0).
\end{equation}
These results arise directly from eqs.\ \ref{eq:165}, conjoined with the Luttinger theorem $I_{L_{A\sigma}}=0$, and are 
\begin{equation}
\label{eq:167}
n_{t}~=~ \sum_{\sigma} \left[
\theta\left(-\epsilon_{f_{A\sigma}}^{*}\right)~+~
\theta\left(
-\epsilon_{c}+\frac{V^{2}}{\epsilon_{f_{A\sigma}}^{*}} -\tfrac{1}{4}t_{*}^{2}G^{R}_{c}(0)
\right)
\right]
\end{equation}
and 
\begin{equation}
\label{eq:168}
\mutil=
\sum_{\sigma}~\sigma \left[
\theta\left(-\epsilon_{f_{A\sigma}}^{*}\right)+
\theta\left(
-\epsilon_{c}+\frac{V^{2}}{\epsilon_{f_{A\sigma}}^{*}} -\tfrac{1}{4}t_{*}^{2}G^{R}_{c}(0)
\right)
\right]-{\cal{J}}
\end{equation}
(with $\sigma = \pm$ for $\uparrow$$/$$\downarrow$-spins), where
\begin{equation}
\label{eq:169}
{\cal{J}} = \tfrac{1}{4}t_{*}^{2}\tfrac{1}{\pi}\mathrm{Im}\int_{-\infty}^{0}d\w~
\left(G_{c_{A\uparrow}}(\w)-G_{c_{A\downarrow}}(\w)\right)\frac{\partial G_{c}(\w)}{\partial\w}.
\end{equation}
They hold throughout the $n_{t}=1$ or $3$ MI phases, and are exact; their derivation is outlined in
Appendix \ref{section: App1}. Eq.\ \ref{eq:167} in particular is the analogue, for the MI, of its counterpart 
eq.\ \ref{eq:117} for the Fermi liquid phases.


\subsubsection{Renormalised levels $\epsilon_{f_{A\sigma}}^{*}$}
\label{subsubsection:rlevelsTSE}

The spin-dependent renormalised levels $\epsilon_{f_{A\sigma}}^{*}$ are characteristic of the MI phase (just as the
renormalised level $\epsilon_{f}^{*}$, eq.\ \ref{eq:113}, is characteristic of Fermi liquid phases); 
and an important element of eqs.\ \ref{eq:167},\ref{eq:168}  is that they imply strong bounds on the 
$\epsilon_{f_{A\sigma}}^{*}$ throughout the MI phase. 

To see this, let us focus on the $n_{t}=1$ MI. Recalling that $\mutil \in (0,1)$ in the MI,
consider first the situation deep in the MI phase. Specifically, it is helpful to have in mind the regime 
$\eta_{f} \in [0,1)$, where for any interaction $U$ the $n_{t}=1$ MI persists to arbitrarily large values of $\epsilon_{c}$ and the moment $\mutil$ asymptotically approaches $1$ (i.e.\ ${\cal{J}} \ll 1$).
Since $n_{t}=1$, it follows from eq.\ \ref{eq:167} that three of the four step functions must vanish; and since 
$\mutil$ cannot be negative it follows from eq.\ \ref{eq:168} that both step functions with $\sigma =\downarrow$ 
must vanish. In other words, the only consistent possibilities are either $\theta (-\epsilon_{f_{A\uparrow}}^{*})=1$ or
$\theta (-\epsilon_{c} +V^{2}/\epsilon_{f_{A\uparrow}}^{*} -\tfrac{1}{4}t_{*}^{2}G_{c}(0)) =1$, with all other step functions vanishing. From this it follows that the $\sigma =\uparrow$ renormalised level necessarily lies above the Fermi level,
\begin{equation}
\label{eq:170}
\epsilon_{f_{A\downarrow}}^{*}~ >~0
\end{equation}
(together with $\epsilon_{c}+\tfrac{1}{4}t_{*}^{2}G_{c}(0) >0$), and that
\begin{equation}
\label{eq:171}
\epsilon_{f_{A\uparrow}}^{*}~ < ~ \epsilon_{f_{A\downarrow}}^{*}.
\end{equation}
But since three of the four step functions must vanish because $n_{t}=1$, it follows that
$\epsilon_{f_{A\downarrow}}^{*}$ cannot change sign throughout the MI phase; otherwise (from eq.\ \ref{eq:168}) $\mutil$ would decrease by $2$ (and thus contradict $\mutil >0$). Eqs.\ \ref{eq:170},\ref{eq:171} thus hold throughout the 
\emph{entire} $n_{t}=1$ MI phase (we illustrate them in sec.\ \ref{subsection:renormnormal}, see  fig.\ \ref{fig:fig10}).


\subsubsection{Asymptotic behaviour of the local moment}
\label{subsubsection:momentasympt}

We turn now to the leading corrections to the  local moment (in the regime $\eta_{f}\in [0,1)$, as above), embodied in 
${\cal{J}}$ (eq.\ \ref{eq:169}), such that $\mutil = 1-{\cal{J}}$. 
Note first that there are two distinct decoupled limits where the moment is fully saturated/spin-polarised:
(a) $V=0$ (but $t_{*} \neq 0$), where the $f$-levels decouple completely from the conduction band
(that ${\cal{J}}$ vanishes here is seen directly from eq.\ \ref{eq:169} on noting that 
$G_{c_{A\uparrow}}(\w) =G_{c_{A\downarrow}}(\w)$ for $V=0$, eq.\ \ref{eq:163a});
(b) $t_{*} =0$ (but $V \neq 0$), the `atomic limit' where sites decouple from each other (with ${\cal{J}}=0$ 
self-evident from eq.\ \ref{eq:169}). As outlined in Appendix \ref{section: App2}, 
${\cal{J}}$ can be obtained exactly to leading order in $V$ (${\cal{O}}(V^{2})$),
but without constraint on either  $\epsilon_{c}/t_{*}$ or $\epsilon_{f}/t_{*}= -\tfrac{U}{2t_{*}}(1-\eta_{f})$.
Denoting by $\tilde{\Delta}$ the dimensionless charge transfer energy, 
\begin{subequations}
\label{eq:172}
\begin{align}
\tilde{\Delta} ~=&~ \left(\epsilon_{c}-\epsilon_{f}\right)/t_{*}
\label{eq:172a}
\\
=&~\tfrac{1}{t_{*}}\left[\epsilon_{c} +\tfrac{1}{2}U(1-\eta_{f})\right],
\label{eq:172b}
\end{align}
\end{subequations}
the local moment is given asymptotically by 
\begin{equation}
\label{eq:173}
\mutil = 1 -2\left(\frac{V}{t_{*}}\right)^{2} \left[Y(\tilde{\Delta})\right]^{2}
\left[
 \left( 1-\frac{1}{\tilde{\Delta}^{2}}
\right)^{-\tfrac{1}{2}}
-1
\right]
\end{equation}
where
\begin{equation}
\label{eq:174}
Y(\tilde{\Delta})~=~ -\tilde{\Delta} ~+~\sqrt{\tilde{\Delta}^{2}-1}~~~~
\overset{\tilde{\Delta} \gg 1}{\sim}~-\frac{1}{2\tilde{\Delta}}~.
\end{equation}
Aside from the explicit factor of $(V/t_{*})^{2}$, note that $\mutil$ is a function solely of the charge transfer energy 
$\tilde{\Delta}$ eq.\ \ref{eq:172} (itself dependent on $\epsilon_{c}/t_{*}$, $U/t_{*}$ and $\eta_{f}$).

Recall (sec.\ \ref{subsection: Phasediagtake2}) that the ZSA classification~\cite{Zaanen-S-A1985}  divides Mott insulators into Mott-Hubbard (MH) insulators for $\Delta \gtrsim U$ and charge-transfer (CT) insulators for $\Delta \lesssim U$, i.e.\ 
(eq.\ \ref{eq:172b})
\begin{equation}
\label{eq:175}
\epsilon_{c}~\gtrless~ \tfrac{1}{2}U (1+\eta_{f}) ~~~~~~:~\mathrm{MH/CT}.
\end{equation}
Well into the MH regime ($\epsilon_{c} \gg U$), $\tilde{\Delta} \sim \epsilon_{c}/t_{*}$ from eq.\ \ref{eq:172b}, 
with $\epsilon_{c}/t_{*} \gg 1$ required for the Mott transition to occur here; in other words, $\tilde{\Delta} \gg 1$.
Likewise, sufficiently far into the CT regime  ($\epsilon_{c} \ll U$), $\tilde{\Delta} \sim  \tfrac{U}{2t_{*}} (1-\eta_{f})$ 
from eq.\ \ref{eq:172b}, with $U/t_{*} \gg 1$ for the Mott transition to occur in this case; again,
 $\tilde{\Delta} \gg 1$. Since $\tilde{\Delta} \gg 1$ in either case, the large-$\tilde{\Delta}$ asymptotic behaviour of 
eq. \ref{eq:173} dominates, \emph{viz}
\begin{equation}
\label{eq:176}
\mutil~\overset{\tilde{\Delta} \gg 1}{\sim}~ 1- \frac{1}{4}\left(\frac{V}{t_{*}}\right)^{2} \frac{1}{\tilde{\Delta}^{4}}~=~
1- \frac{1}{4}~\frac{V^{2}t_{*}^{2}}{\left[\epsilon_{c} +\tfrac{1}{2}U(1-\eta_{f})\right]^{4}}.
\end{equation}
The high(4$^{\mathrm{th}}$)-order character of this leading perturbative correction is the natural reason why 
$\mutil$ is so well-developed throughout the entire MI phase, see e.g.\ fig.\ \ref{fig:fig9}
[for the standard one-band Hubbard model by contrast, the corresponding leading 
correction is second order in $t_{*}/U \ll 1$, \emph{viz}~\cite{DELMRG2016} $\mutil -1 = (t_{*}/2U)^{2}$, 
with the moment accordingly less fully developed].

The results eqs.\ \ref{eq:173},\ref{eq:176} are compared to NRG calculations of $\mutil$ in fig.\ \ref{fig:fig9}, 
and the leading asymptotic behaviour eq.\ \ref{eq:176} seen clearly to emerge  with increasing $\epsilon_{c}/t_{*}$. 
The `full' asymptotic form eq.\ \ref{eq:173} is also seen to capture remarkably well the NRG results over the full 
$\epsilon_{c}/t_{*}$-range in the Mott insulator. Indeed one might turn the latter comment on its head,  to note the 
rather impressive accuracy with which NRG can capture deviations from saturation on the order of $10^{-4}$.


\subsection{Renormalised levels $\epsilon_{f}^{*}$, and Luttinger theorem}
\label{subsection:renormnormal}

In considering the MI phase, we have naturally focused on the renormalised levels $\epsilon_{f_{A\sigma}}^{*}$ 
given in terms of the two-self-energies. As in the metallic Fermi liquid phase, however, one can also consider the 
($\sigma$-independent) renormalised levels defined in terms of the single self-energy and given by eq.\ \ref{eq:113}, 
$\epsilon_{f}^{*}= \epsilon_{f}+\Sigma_{f}^{R}(0)$.

As has been exploited throughout, the ph-symmetry $H(\epsilon_{c},\eta_{f}) \equiv H(-\epsilon_{c},-\eta_{f})$ enables consideration solely of $\eta_{f} \geq 0$. Here, as we have seen, both $n_{t}=1$ and $n_{t}=3$ Mott insulators arise (now unrelated under a ph-transformation); with the $n_{t}=1$ MI occurring only for $\epsilon_{c}/t_{*} >1$, and the 
$n_{t}=3$ MI only for $\epsilon_{c}/t_{*} <-1$. In addition, we have the general condition $|-\epsilon_{c}+V^{2}/\epsilon_{f}^{*}| >t_{*}$ (eq.\ \ref{eq:116}) for an insulating phase to occur.
The condition $\epsilon_{c}>t_{*}$, conjoined with eq.\ \ref{eq:116}, implies directly that 
$\epsilon_{f}^{*} >0$ for the $n_{t}=1$ MI, with the following bounds on the renormalised levels
\begin{equation}
\label{eq:177}
0 < \epsilon_{f}^{*} < \frac{V^{2}}{\epsilon_{c}+t_{*}} ~~~~\mathrm{or}~~~~
0 < \frac{V^{2}}{\epsilon_{c}-t_{*}} <\epsilon_{f}^{*}
~~~:n_{t}=1
\end{equation}
according to whether $-\epsilon_{c}+V^{2}/\epsilon_{f}^{*}>t_{*}$ or $<-t_{*}$ respectively.
Likewise, for the $n_{t}=3$ MI the condition $\epsilon_{c} <-t_{*}$ together  with eq.\ \ref{eq:116} implies  
$\epsilon_{f}^{*} <0$,  with corresponding bounds $\epsilon_{f}^{*} < V^{2}/(\epsilon_{c}+t_{*})<0$
or $V^{2}/(\epsilon_{c}-t_{*}) <\epsilon_{f}^{*}<0$.

\begin{figure}
\includegraphics{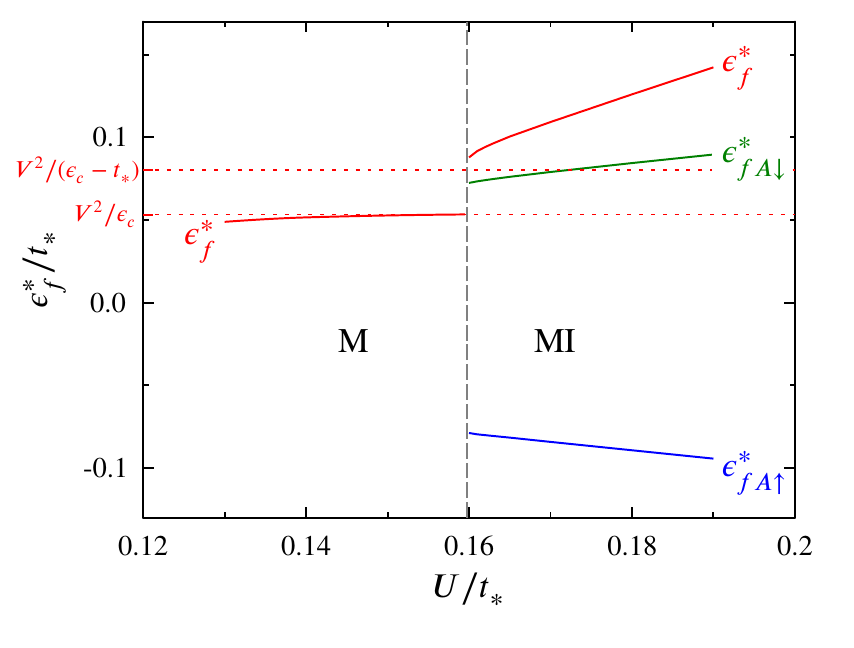}
\caption{\label{fig:fig10} 
NRG results for renormalised levels, shown \emph{vs} $U/t_{*}$ for fixed $\epsilon_{c}/t_{*}=3$, $V/t_{*}=0.4$ and 
$\eta_{f}=0$ (where the critical $U_{c}/t_{*} \simeq 0.160$). On approaching the $n_{t}=1$ Mott transition from the
metal, $\epsilon_{f}^{*}$ indeed tends to the limit $\epsilon_{f}^{*}= V^{2}/\epsilon_{c}$ (eq.\ \ref{eq:121}), and then jumps discontinuously on entering the MI, while remaining $>0$ (eq.\ \ref{eq:177}).
Approaching the transition from the MI, the renormalised levels $\epsilon_{f_{A\downarrow}}^{*}$ and
$\epsilon_{f_{A\uparrow}}^{*}$ remain finite, with $\epsilon_{f_{A\downarrow}}^{*}>0$ and 
$\epsilon_{f_{A\uparrow}}^{*}<\epsilon_{f_{A\downarrow}}^{*}$ as required (eqs.\ \ref{eq:170},\ref{eq:171}).
}
\end{figure}

As shown in sec.\ \ref{subsubsection:renormlevels1},
on approaching either the $n_{t}=1$ or $n_{t}=3$ Mott transition from the \emph{metallic} phase, the renormalised level 
$\epsilon_{f}^{*}$ approaches the limiting value $\epsilon_{f}^{*} = V^{2}/\epsilon_{c}$ (see eq.\ \ref{eq:121}). From 
the above it thus follows that, while $\mathrm{sgn}(\epsilon_{f}^{*})$ remains fixed as the Mott transition is crossed, 
a discontinuous jump occurs in the renormalised level $\epsilon_{f}^{*}$.
NRG results exemplifying this behaviour are given in fig.\ \ref{fig:fig10}, which also shows the calculated renormalised levels $\epsilon_{f_{A\downarrow}}^{*}$, $\epsilon_{f_{A\uparrow}}^{*}$ (sec.\ \ref{subsubsection:rlevelsTSE}).

\subsubsection{Luttinger theorem for $I_{L}$}
\label{section:Luttint}

Luttinger's theorem $I_{L}=0$ holds throughout the Fermi liquid phases, with the usual Luttinger integral
$I_{L}$ given by eq.\ \ref{eq:118}. It does not  however hold in the MI, so  what can be deduced about $I_{L}$ in this phase?

  Consideration of the two-self-energies, and associated renormalised levels $\epsilon_{f_{A\sigma}}^{*}$, is 
mandatory to say anything about the local moment $\mutil$ (including its mere existence). But that is not the case for the total charge $n_{t}$. In this case one can repeat the analysis of sec.\ \ref{subsection:C&MMI}, 
working with the single self-energy $\Sigma_{f}(\w)$ instead of the 
$\{\Sigma_{f_{A\sigma}}(\w)\}$ -- but without assuming a vanishing Luttinger integral $I_{L}$.
The resultant analogue of eq.\ \ref{eq:167} for $n_{t}$ is then
\begin{equation}
\label{eq:178}
\tfrac{1}{2}n_{t}
~=~\Big[\theta (-\epsilon_{f}^{*})+
\theta\Big(-\epsilon_{c} +\frac{V^{2}}{\epsilon_{f}^{*}} -\tfrac{1}{4}t_{*}^{2}G_{c}(0)\Big)\Big] 
-\tfrac{1}{\pi}I_{L}
\end{equation}
where $n_{t}=1$ or $3$ for the MI phase. $G_{c}(0)$ itself is moreover a function of $\epsilon_{f}^{*}$:
eq.\ \ref{eq:18a} gives $G_{c}(0)= [-\epsilon_{c}-\tfrac{1}{4}t_{*}^{2}G_{c}(0) +V^{2}/\epsilon_{f}^{*}]^{-1}$, 
and hence a quadratic for $G_{c}(0)$, from which it is easily shown that the sign of 
$-\epsilon_{c} +V^{2}/\epsilon_{f}^{*}-\tfrac{1}{4}t_{*}^{2}G_{c}(0)$ is necessarily that of 
$-\epsilon_{c}+V^{2}/\epsilon_{f}^{*}$. Eq.\ \ref{eq:178} then gives
\begin{equation}
\label{eq:179}
\tfrac{1}{\pi}I_{L}~=~\Big[\theta (-\epsilon_{f}^{*})+
\theta\Big(-\epsilon_{c} +\frac{V^{2}}{\epsilon_{f}^{*}}\Big)\Big] -\tfrac{1}{2}n_{t}~.
\end{equation}
Under a ph-transformation where $(\epsilon_{c},\eta_{f})\rightarrow(-\epsilon_{c},-\eta_{f})$, 
the renormalised level changes sign, $\epsilon_{f}^{*} \rightarrow -\epsilon_{f}^{*}$ (eq.\ \ref{eq:114}), 
and the total charge $n_{t} \rightarrow 4-n_{t}$ (eq.\ \ref{eq:110}).
From eq.\ \ref{eq:179}, the Luttinger integral thus inverts under a ph-transformation, 
$I_{L} \rightarrow -I_{L}$. 
As usual, we can thus focus on the $n_{t}=1$ or $3$ MIs arising for $\eta_{f} \geq 0$.
Here, as shown in sec.\ \ref{subsection:renormnormal}, 
$\epsilon_{f}^{*} >0$ for the $n_{t}=1$ MI, while 
$\epsilon_{f}^{*} <0$ for the $n_{t}=3$ MI.
From eq.\ \ref{eq:179} it follows that, for both the $n_{t}=1$ and $3$ MIs, the Luttinger integral is given by
\begin{subequations}
\label{eq:180}
\begin{align}
I_{L}~=&~ +\frac{\pi}{2} ~~~~~~:~-\epsilon_{c}+\frac{V^{2}}{\epsilon_{f}^{*}} >t_{*}
\\
=&~-\frac{\pi}{2} ~~~~~~:~-\epsilon_{c}+\frac{V^{2}}{\epsilon_{f}^{*}} <-t_{*}
\end{align}
\end{subequations}
(noting also the bounds eq.\ \ref{eq:116} required for an insulator).

The magnitude of the Luttinger integral is thus $\pi/2$ throughout the MI phases, \emph{independent} of  
interaction strength (or indeed of any underlying model parameters, provided only the system is a MI).
This generalises to the non-Fermi liquid MI the familiar Luttinger theorem $I_{L}=0$ applicable to the Fermi liquid phases; 
and, since it encompasses the $V=0$ limit where the $f$-levels decouple completely from the 
$c$-band, suggests it reflects perturbative continuity to that limit (akin to the fact that $I_{L}=0$ for Fermi liquid phases reflects adiabatic continuity to the non-interacting limit $U=0$). The same result is moreover 
found for the local moment phases of a wide range of quantum impurity models \emph{per se},~\cite{LTG2014}
as well as for the one-band NN Hubbard model within DMFT;~\cite{DELMRG2016} suggesting its ubiquity as a 
hallmark of the locally degenerate ground states arising in both the local moment phases of quantum impurity models and, 
relatedly, the MI phases of lattice-fermion models within DMFT.

\begin{acknowledgments}
We are grateful to the EPSRC for financial support under grants EP/L015722/1 and EP/N01930X/1; and
the work is compliant with EPSRC Open Data requirements. We thank A. Mitchell for helpful discussions about several 
aspects of this work.
One of us (DEL) also expresses his warm thanks for the hospitality of the Physics Department, Indian Institute of Science, Bangalore, where part of this work was completed; and to H. R. Krishnamurthy, T. V. Ramakrishnan and N. S. Vidhyadhiraja for
stimulating discussions. DEL would also like to acknowledge, with gratitude, numerous enlightening discussions 
about correlated electron systems over many years, with his late friend and sometime collaborator, Thomas Pruschke.

\end{acknowledgments}


\appendix  

\section{Total charge and local moment in Mott insulator phase}
\label{section: App1}
Here we outline the derivation of eqs.\ \ref{eq:167},\ref{eq:168} for the total site charge $n_{t}$ and local moment
$\mutil$ in the MI phase. The propagators $G_{c_{A\sigma}}(\w)$ and $G_{f_{A\sigma}}(\w)$ are given by eqs.\ \ref{eq:163}; from which follows an identity relating the two:
\begin{equation}
\label{eq:A1}
\begin{split}
&G_{f_{A\sigma}}(\w)~=~
\\
&~\frac{1}{[\w^{+}-\epsilon_{f} -\Sigma_{f_{A\sigma}}(\w)]}
\left[
1+ \frac{V^{2}}{[\w^{+}-\epsilon_{f} -\Sigma_{f_{A\sigma}}(\w)]}G_{c_{A\sigma}}(\w)
\right]
\end{split}
\end{equation}
Eqs.\ \ref{eq:163a} and\ref{eq:A1} thus give 
\begin{equation}
\label{eq:A2}
\begin{split}
\sum_{\sigma} l_{\sigma} (& G_{c_{A\sigma}}(\w) +G_{f_{A\sigma}}(\w)) =
\sum_{\sigma} l_{\sigma} \left[\w^{+}-\epsilon_{f}-\Sigma_{f_{A\sigma}}(\w) \right]^{-1}
\\
+& \sum_{\sigma} l_{\sigma} \left(
1+\frac{V^{2}}{[\w^{+}-\epsilon_{f}-\Sigma_{f_{A\sigma}}(\w)]^{2}}
\right) G_{c_{A\sigma}}(\w)
\end{split}
\end{equation}
where $l_{\sigma}$ here denotes either  $1$ or $\sigma$ (and with $\sigma =\pm$ for $\uparrow$$/$$\downarrow$ spins, as usual).

The following is simply an identity,
\begin{equation}
\begin{split}
&\left(1+\frac{V^{2}}{[\w^{+}-\epsilon_{f}-\Sigma_{f_{A\sigma}}(\w)]^{2}}\right)G_{c_{A\sigma}}(\w) ~=~
\\
&
\frac{\partial}{\partial\w}\ln \left(
\w^{+}-\epsilon_{c} -\frac{V^{2}}{\w^{+}-\epsilon_{f}-\Sigma_{f_{A\sigma}}(\w)} -\tfrac{1}{4}t_{*}^{2}G_{c}(\w)\right) 
\\ &+
\left[ 
G_{f_{A\sigma}}(\w) - \tfrac{1}{[\w^{+}-\epsilon_{f} -\Sigma_{f_{A\sigma}}(\w)]}
\right]\frac{\partial\Sigma_{f_{A\sigma}}(\w)}{\partial\w} 
\\
&+G_{c_{A\sigma}}(\w)\tfrac{1}{4}t_{*}^{2}\frac{\partial G_{c}(\w)}{\partial\w}
\label{eq:A3}
\end{split}
\end{equation}
where the left side of eq.\ \ref{eq:A3} appears in eq.\ \ref{eq:A2} (and eq.\ \ref{eq:A1} has been used).
With it, eq.\ \ref{eq:A2} gives
\begin{equation}
\label{eq:A4}
\begin{split}
\tfrac{(-1)}{\pi}&\mathrm{Im}\sum_{\sigma} l_{\sigma} ~\int_{-\infty}^{0}d\w ~( G_{c_{A\sigma}}(\w) +G_{f_{A\sigma}}(\w)) 
~=~
\\
& \tfrac{(-1)}{\pi}\mathrm{Im}\sum_{\sigma} l_{\sigma} \int_{-\infty}^{0}d\w\Bigg[
\frac{\partial}{\partial\w} \ln\left(\w^{+}-\epsilon_{f}-\Sigma_{f_{A\sigma}}(\w)\right)
\\
+&\frac{\partial}{\partial\w}\ln \left(
\w^{+}-\epsilon_{c} -\frac{V^{2}}{\w^{+}-\epsilon_{f}-\Sigma_{f_{A\sigma}}(\w)} -\tfrac{1}{4}t_{*}^{2}G_{c}(\w)\right)
\\
+& G_{c_{A\sigma}}(\w) \tfrac{1}{4}t_{*}^{2}\frac{\partial G_{c}(\w)}{\partial\w}\Bigg]
\end{split}
\end{equation}
on using the Luttinger theorem eq.\ \ref{eq:164}, $I_{L_{A\sigma}}=0$ (which holds for either spin $\sigma$ throughout the degenerate MI phase). With $l_{\sigma}=1$, eq.\ \ref{eq:A4} gives $n_{t}$, and with $l_{\sigma}= \sigma$ it gives $\mutil$,
see  eqs.\ \ref{eq:165}.

The two $\ln$-derivative terms in eq.\ \ref{eq:A4} can be evaluated explicitly, using $\Sigma_{f_{A\sigma}}^{I}(0) =0$ and
$\tfrac{(-1)}{\pi}\mathrm{Im}G_{c}(\w =0) =D_{c}(0) =0$, as appropriate to the Mott insulator. With this,
eq.\ \ref{eq:A4} becomes
\begin{equation}
\label{eq:A5}
\begin{split}
\tfrac{(-1)}{\pi}&\mathrm{Im}\sum_{\sigma} l_{\sigma} ~\int_{-\infty}^{0}d\w ~( G_{c_{A\sigma}}(\w) +G_{f_{A\sigma}}(\w)) 
~=~
\\
&\sum_{\sigma} l_{\sigma} \Bigg(
\theta \left(-\epsilon_{f_{A\sigma}}^{*}\right)
+ \theta \Big(
-\epsilon_{c} +\frac{V^{2}}{\epsilon_{f_{A\sigma}}^{*}} -\tfrac{1}{4}t_{*}^{2}G_{c}(0)
\Big)
\\
& ~~~~+\tfrac{(-1)}{\pi}\mathrm{Im}\int_{-\infty}^{0}d\w~G_{c_{A\sigma}}(\w) \tfrac{1}{4}t_{*}^{2}
\frac{\partial G_{c}(\w)}{\partial\w}\Bigg)
\end{split}
\end{equation}
where $\epsilon_{f_{A\sigma}}^{*}$ are the renormalised levels, eq.\ \ref{eq:166}.
With $l_{\sigma}=\sigma$, eq.\ \ref{eq:A5} gives directly eqs.\ \ref{eq:168},\ref{eq:169} for the moment $\mutil$.

Now consider $l_{\sigma}\equiv 1$, such that the left side of eq.\ \ref{eq:A5} gives $n_{t}$ (eq.\ \ref{eq:165a}). Since
$ \sum_{\sigma} G_{c_{A\sigma}}(\w) = 2G_{c}(\w)$ (eq.\ \ref{eq:162}), 
\begin{equation}
\begin{split}
\sum_{\sigma}\tfrac{(-1)}{\pi}&\mathrm{Im}\int_{-\infty}^{0}d\w~G_{c_{A\sigma}}(\w) \tfrac{1}{4}t_{*}^{2}
\frac{\partial G_{c}(\w)}{\partial\w}
\\
=&~ \tfrac{1}{4}t_{*}^{2} \tfrac{(-1)}{\pi}\mathrm{Im} \left(
\left[ G_{c}(0)\right]^{2}\right)
~=~0
\end{split}
\label{eq:A6}
\end{equation}
(as $D_{c}(0)=0$ and $G_{c}(\w =-\infty) =0$).
With this, eq.\ \ref{eq:A5} yields directly eq.\ \ref{eq:167} for $n_{t}$.


\section{Asymptotic behaviour of $\mutil$.}
\label{section: App2}

We outline the origins of eqs.\ \ref{eq:173},\ref{eq:176} for the asymptotic behaviour of the local moment 
$\mutil = 1-{\cal{J}}$ (with $\cal{J}$ given by eq.\ \ref{eq:169}), in the regime $\eta_{f} \in [0,1)$ considered.

Eq.\ \ref{eq:163a} for $G_{c_{A\sigma}}(\w)$ may be expanded to leading order in $V^{2}$, to give:
\begin{equation}
\label{eq:B1}
\begin{split}
&G_{c_{A\uparrow}}(\w)-G_{c_{A\downarrow}}(\w) =
V^{2}~\frac{1}{[\w^{+}-\epsilon_{c}-\tfrac{1}{4}t_{*}^{2}G_{c}(\w)]^{2}}~
\\
&
\times \left[
\frac{1}{[\w^{+}-\epsilon_{f}-\Sigma_{f_{A\uparrow}}(\w)]}-
\frac{1}{[\w^{+}-\epsilon_{f}-\Sigma_{f_{A\downarrow}}(\w)]}
\right]
\end{split}
\end{equation}
Given the explicit $V^{2}$ factor here, to obtain $\cal{J}$ to leading $V^{2}$ order thus requires
$\Sigma_{f_{A\sigma}}(\w)$ and $G_{c}(\w)$ for $V=0$. This is simple, because for $V=0$ the $f$- and $c$-levels decouple.
Since the $f$-levels are then free, the $\Sigma_{f_{A\sigma}}(\w)$ are thus given by~\cite{LTG2014}
$\Sigma_{f_{A\uparrow}}(\w)=0$ and $\Sigma_{f_{A\downarrow}}(\w) =U$; 
while $[\w^{+}-\epsilon_{c}-\tfrac{1}{4}t_{*}^{2}G_{c}(\w)]^{-1}$ for $V=0$ reduces simply to
$g_{c}(\w)$  (see eq.\ \ref{eq:126}). Hence, writing $\epsilon_{f} = -\tfrac{1}{2}U(1-\eta_{f})$ and
$\epsilon_{f}+U = \tfrac{1}{2}U(1+\eta_{f})$, 
\begin{equation}
\label{eq:B2}
\begin{split}
G_{c_{A\uparrow}}&(\w)~-~G_{cA_{\downarrow}}(\w) ~=~V^{2}~[g_{c}(\w)]^{2}
\\
&\times
\left[
\frac{1}{\w^{+}+\tfrac{1}{2}U(1-\eta_{f})} -\frac{1}{\w^{+}-\tfrac{1}{2}U(1+\eta_{f})}
\right]
\end{split}
\end{equation}
is exact to leading order in $V^{2}$, but for \emph{any} $\epsilon_{c}/t_{*}$, $U$ and $\eta_{f}$ throughout the MI phase.

Recall that $\epsilon_{c}/t_{*} >1$ is required for the $n_{t}=1$ MI. In consequence, $g_{c}(\w)$ is pure real for all
$\w \leq 0$ ($\equiv E_{F}$), and given by (eq.\ \ref{eq:126})
\begin{equation}
\label{eq:B3}
\tfrac{1}{2}t_{*} g_{c}(\w) ~=~\frac{(\w -\epsilon_{c})}{t_{*}} +
\left[
\left(\frac{\w -\epsilon_{c}}{t_{*}}\right)^{2} -1
\right]^{1/2}~.
\end{equation}
Employing eq.\ \ref{eq:B2} in eq.\ \ref{eq:169} for $\cal{J}$ and performing the $\w$-integration, gives
\begin{equation}
\label{eq:B4}
{\cal{J}}=-\tfrac{1}{4}V^{2}t_{*}^{2}~
\left(
[g_{c}(\w)]^{2}~\frac{\partial g_{c}(\w)}{\partial \w}
\right)_{\w =-\tfrac{1}{2}U(1-\eta_{f})}.
\end{equation}
Evaluation of eq.\ \ref{eq:B4} using eq.\ \ref{eq:B3} then gives directly eq.\ \ref{eq:173}
for $\mutil = 1- {\cal{J}}$.


\bibliography{paper}

\begin{thebibliography}{60}
\bibitem [{fne()}]{fnepsf*=0}%
  \BibitemOpen
  \href@noop {} {}\bibinfo {note} {The case $\epsilon_{f}^{*}=0$ in eq.\
  \ref{eq:117} is to be understood as the limit $\epsilon_{f}^{*}\rightarrow 0$
  (giving $n_{t}=2$ whether $\epsilon_{f}^{*}\rightarrow 0+$ or
  $0-$).}\BibitemShut {Stop}%
\bibitem [{KIM()}]{KIMboundary}%
  \BibitemOpen
  \href@noop {} {}\bibinfo {note} {They further imply that as $U \rightarrow
  \infty$ (where $\epsilon_{c}\rightarrow 0$ for the KI/M transition), the
  renormalised level changes discontinuously across $\epsilon_{c}=0$, from
  $\epsilon_{f}^{*}=-V^{2}/t_{*}$ for $\epsilon_{c}=0-$ to $+V^{2}/t_{*}$ for
  $\epsilon_{c}=0+$.}\BibitemShut {Stop}%
\bibitem [{ome()}]{omegaLdef}%
  \BibitemOpen
  \href@noop {} {}\bibinfo {note} {$\omega_{_{\mathrm{L}}}$ is defined in
  practice as the right ($\w >0$) half-width at half-maximum in
  $D_{c}(\w)$.}\BibitemShut {Stop}%
\end{thebibliography}

\end{document}